# Multiple and Fast: The Accretion of Ordinary Chondrite Parent Bodies

P. Vernazza[1], B. Zanda[2,6], R. P. Binzel[3,7], T. Hiroi[4], F. E. DeMeo[3], M. Birlan[5], R. Hewins[2,6], L. Ricci[8], P. Barge[1], M. Lockhart[3]

[1]Aix Marseille Université, CNRS, LAM (Laboratoire d'Astrophysique de Marseille) UMR 7326, 13388, Marseille, France

[2]Institut de Minéralogie, de Physique des Matériaux, et de Cosmochimie (IMPMC) - Sorbonne Universités - Muséum National d'Histoire Naturelle, UPMC Université Paris 06, UMR CNRS 7590, IRD UMR 206, 61 rue Buffon, F-75005 Paris, France.

[3]Department of Earth, Atmospheric, and Planetary Sciences, Massachusetts Institute of Technology, Cambridge, MA 02139, USA

[4]Department of Geological Sciences, Brown University, Providence, RI 02912, USA

[5]IMCCE, Observatoire de Paris, 77 Av. Denfert Rochereau, 75014 Paris Cedex, France

[6]Department of Earth and Planetary Sciences, Rutgers University, Piscataway NJ 08854, USA

[7]Chercheur Associé, IMCCE, Observatoire de Paris, 77 Av. Denfert Rochereau, 75014 Paris Cedex, France

[8]California Institute of Technology, MC 249-17, Pasadena, CA, 91125, USA

**Abstract**

Although petrologic, chemical and isotopic studies of ordinary chondrites and meteorites in general have largely helped establish a chronology of the earliest events of planetesimal formation and their evolution, there are several questions that cannot be resolved via laboratory measurements and/or experiments only. Here we propose rationale for several new constraints on the formation and evolution of ordinary chondrite parent bodies (and by extension most planetesimals) from newly available spectral measurements and mineralogical analysis of main belt S-type asteroids (83 objects) and unequilibrated ordinary chondrite meteorites (53 samples). Based on the latter, we suggest spectral data may be used to distinguish whether an ordinary chondrite was formed near the surface or in the interior of its parent body. If these constraints are correct, the suggested implications include that: i) large groups of compositionally similar asteroids are a natural outcome of planetesimal formation and, consequently, meteorites within a given class can originate from multiple parent bodies; ii) the surfaces of large (up to ~200km) S-type main-belt asteroids expose mostly the interiors of the primordial bodies, a likely consequence of impacts by small asteroids (D<10km) in the early solar system (Ciesla et al. 2013); iii) the duration of accretion of the H chondrite parent bodies was likely short (instantaneous or in less then ~$10^5$ yr but certainly not as long as 1 Myr); iv) LL-like bodies formed closer to the Sun than H-like bodies, a possible consequence of radial mixing and size sorting of chondrules in the protoplanetary disk prior to accretion.

# 1) Introduction

Ordinary chondrite meteorites (OCs) are by far the most abundant meteorites (80% of all falls; Hutchison 2004). They are subdivided into three groups (H, L and LL) based on variations in bulk composition, such as molecular ratios [FeO ∕ (FeO+MgO)] in olivine and pyroxene (Mason 1963, Keil & Fredriksson 1964) and the ratio of metallic Fe to total Fe (Dodd et al. 1967). Their study along with that of other chondrite classes has provided numerous constraints on the formation and early evolution of the solar system, including **a)** the migration processes that occurred in the protoplanetary disk prior to primary accretion (i.e. planetesimal formation) and their associated timescales (Cuzzi et al. 2001, Cuzzi & Weidenschilling 2006), **b)** the post- (and syn-) accretional heating events (Huss et al. 2006, Ghosh et al. 2006), and **c)** the collisional events that occurred since their accretion (Hutchison 2004, Haack et al. 1996 and references therein):

**a)** Unequilibrated ordinary chondrites (UOCs), which represent ~15 % of all OCs (Hutchison 2004), are the most primitive OCs (see Appendix Fig. A1). Several thermometers indicate that a number of UOCs did not experience metamorphic temperatures (Hutchison 2004) > ~370 ºC and that all were < ~600 ºC. As such, UOCs, give us the best indication of the initial material (a sort of snapshot of the protoplanetary disk) from which their parent asteroids accreted, as well as key constraints on the events that occurred in the protoplanetary disk prior to primary accretion.

UOCs are mainly aggregates of high-temperature (> 1600 ºC) components, including chondrules (~80% of the volume) which are millimeter-sized silicate spherules that formed through rapid melting and cooling of precursor material, via a still elusive mechanism (Connolly and Desch 2004; Jacquet et al. 2012), as well as metal and sulfide grains. All these components are set in an opaque, fine-grained interchondrule matrix (10-15 vol% of the rock) (Hutchison 2004), which is believed to have always remained at low-temperature (< 200 ºC) in the disk.

As of today, the most likely explanation for the simultaneous presence in ordinary chondrites of low-temperature and high-temperature components believed to have formed respectively far from and close to the Sun is that radial mixing was extensive in the solar nebula and thus played a prominent role in shaping the composition of these chondrites and that of planetesimals overall. Indeed, evidence of the simultaneous presence of low-temperature and high-temperature components is also observed in other chondrite classes (Hutchison 2004 and references therein) and in comets (Kelley & Wooden 2009 and references therein).

Further evidence of the importance of radial mixing in all classes of chondrites, and ordinary chondrites in particular, is provided by i) the 'universal shape' of the size distribution of chondrules that is centered at different sizes from chondrite class to chondrite class (Grossman et al. 1989 and references therein, Cuzzi et al. 2001, 2008), indicative of size-sorting and ii) the simultaneous presence in a given chondrite of various proportions of two different types of chondrules (reduced type I and oxidized type II chondrules – made in different environments, Zanda et al. 2006; see Appendix Fig. A1).

**b)** Textural variations and corresponding mineral and chemical trends indicate that differing degrees of thermal metamorphism (heating) took place within each chondrite group (Van Schmus & Wood 1967; Dunn et al. 2010b). Based on these variations, a petrologic classification scheme (Van Schmus & Wood 1967) for ordinary chondrites was developed (H, L and LL groups are further subdivided into fours groups – from 3 to 6), which consisted in distinguishing the less metamorphosed chondrites (type 3, called unequilibrated ordinary chondrites: UOCs) from chondrites that have undergone higher degrees of thermal metamorphism (type 4 to 6, called equilibrated ordinary chondrites: EOCs; see Huss et al. 2006 for a review).

There are several heat sources that have been proposed to explain the thermal evolution (including metamorphism and/or differentiation) of planetesimals and thus the existence of the different petrologic types (3 to 6) within each OC class (see McSween et al. 2002, Ghosh et al. 2006, Sahijpal et al. 2007 for reviews). These include the decay of short-lived radioactive nuclides (Urey 1955), the decay of long-lived radioactive elements only (Yomogida & Matsui 1984), electromagnetic induction heating (Sonnett et al. 1968) and impact heating (e.g., Rubin 1995, 2003, 2004). Recent work has shown that neither electromagnetic induction nor impacts alone can explain the thermal processing of planetesimals [see Marsh et al. (2006) concerning electromagnetic induction and Keil (1997) and Ciesla et al. (2013) for impact heating]. There is also no coherent scenario that could favor the decay of long-lived radioactive elements as the only heat source. In such a scenario, the different petrologic types would have formed on separate parent bodies (Yomogida & Matsui 1984), which is in contradiction with the cosmic ray exposure (CRE) ages of individual OCs. The latter show that for some classes (mainly H chondrites), several petrologic types (mainly types 4 and 5 in the H case) (Marti & Graf 1992; Graf & Marti 1994, Graf & Marti 1995) show similar peaks in their CRE age distribution, thus suggesting that the different petrologic types do originate from the same parent body.

The decline in the role attributed to impacts (either early or late ones, see Ciesla et al. 2013), induction heating and the decay of long-lived radioactive elements favors the decay of short-lived radioactive nuclides as the primary heat source for the metamorphism and/or the differentiation of planetesimals (Sahijpal et al. 2007). Accordingly, several groups have developed a wide range of thermal models of planetesimals with short-lived radioactive nuclides (mainly $^{26}$Al but also $^{60}$Fe) as the main heat source (e.g., Miyamoto et al. 1981;

Miyamoto 1991; Grimm and McSween 1993; Bennett and McSween 1996; Akridge et al. 1998; Merk et al. 2002; Ghosh et al. 2003; Tachibana & Huss 2003; Trieloff et al. 2003; Bizzarro et al. 2005; Baker et al. 2005; Mostefaoui et al. 2005; Hevey and Sanders 2006; Sahijpal et al. 2007; Harrison & Grimm 2010; Elkins-Tanton et al. 2011; Henke et al. 2012a,b; Neumann et al. 2012, Monnereau et al. 2013). In the context of ordinary chondrite parent bodies, such models predict an onion-shell structure (Trieloff et al. 2003; Ghosh et al. 2006; Henke et al. 2012a,b;2013 and references therein) where all petrologic types formed on the same parent body with type 3 OCs being representative of the crust and type 6 OCs being representative of the core.

**c)** Both the cooling rates and the gas-retention ages of meteorites have provided strong constraints on the collisional events that occurred since accretion:

- The metallographic cooling rates of H chondrites suggest that several early impacts punctured the H chondrite parent body (bodies) to type 6 depths (assuming an onion-shell structure) while it was cooling, causing disturbances in the thermal histories of many H chondrites and leading to surfaces containing rocks that originated at a wide range of depths (Taylor et al. 1987; Scott et al. 2011, 2013, 2014; Krot et al. 2012, Ciesla et al. 2013).

- Argon-isotope gas-retention ages of meteorites show that a major asteroid disruption event affecting the L chondrite parent body occurred in the asteroid belt at about 470 million years ago (Ma) (Haack et al. 1996 and references therein).

Although petrologic, chemical and isotopic studies of OCs and meteorites in general have largely helped establish a chronology of the earliest events of planetesimal formation, there are several questions that cannot be resolved via laboratory measurements and/or experiments only. These include the formation location of the different classes of ordinary chondrites (and meteorites in general); the initial average size of their parent bodies; the amplitude of the bias in our collections with respect to the compositional distribution of OC-like material in the Asteroid Belt; the number of parent bodies for a given meteorite class (it is typically proposed that each meteorite class has only one parent body); the level of radial mixing experienced by parent bodies after their formation; and their accretion timescale. To investigate answers to these questions, we conducted an extensive spectroscopic survey of 83 main belt S-type asteroids (Appendix Table 1, Appendix Fig. 2) and members of 3 S-type families as it was recently established unambiguously (Nakamura et al. 2011) that these asteroids encompass the parent bodies of OCs. In parallel, we also obtained for the first time spectral measurements for a representative number (53) of UOCs (Appendix Table 2) as those were lacking in current databases (e.g., RELAB; http://www.planetary.brown.edu/relab/).

## 2) Telescopic Observations and laboratory measurements

### 2.1 Telescopic Observations

To explore the mineralogical composition of our asteroid sample, we utilized the SpeX instrument (Rayner et al. 2003) over the near-infrared 0.8-2.5 micron range on the NASA Infrared Telescope Facility (IRTF) on Mauna Kea, Hawaii. This wavelength range covers the diagnostic absorption bands (at 1 and 2 μm) due to olivine and pyroxene, which are characteristic for both OCs and S-type asteroids (see A1 for a brief description of the observation protocol and the data reduction procedure). Combining our new near-infrared measurements with available visible wavelength spectra (Bus 1999) gives for the first time a nearly complete spectral database of main-belt S-types with D>60 km (95% or 54/56; accounting for >85% of all S-type mass). Previous near-IR measurements were acquired for 32 main belt S-types (Gaffey et al. 1993) while Deleon et al. (2010) acquired near-IR spectra for 22 main belt S-types and Gietzen et al. (2012) for 5 main belt S-types.

## 2.2 Laboratory measurements

While spectral data for more than 60 equilibrated ordinary chondrites (EOCs) have existed within the RELAB database for a long time (http://www.planetary.brown.edu/relab/), similar data have been quasi inexistent for type 3 UOCs [there were spectra for only 11 UOCs: Suwahib -Buwah (H3.7), Dhajala (H3.8), ALHA77214 (L3.4), Hallingeberg (L3.4), Khohar (L3.6), Mezo-Madaras (L3.7), Hedjaz (L3.7), Bishunpur (LL3.1), Krymka (LL3.1), Chainpur (LL3.4), Parnallee (LL3.6)]. Considering that the range of properties within petrologic type 3 (subtypes 3.0-3.9) is as great as that from types 4-6 (Sears et al. 1980), it is clear that there is a major lack of data within the RELAB spectral database.

In order to reduce the current gap in data between UOCs and EOCs, we obtained a large number of Antarctic samples (http://curator.jsc.nasa.gov/antmet/) with the specific goal of measuring the spectral properties of UOCs. As UOCs are compositionally diverse, we asked for a representative sample (53 meteorites in total) for H, L and LL UOCs (i.e. the 3.0-3.9 continuum needing to be well covered for H, L and LL – See Appendix Table 2). The meteorites were sieved to grain sizes in the 0-45 micron range and spectra were subsequently collected at the RELAB facility (http://www.planetary.brown.edu/relab/) over the 0.3-2.6 micron range. The spectra are available in the RELAB database.

## 3) Compositional analysis of our meteorite and asteroid samples

We performed a detailed comparison of our telescopically measured asteroid spectra (i.e. measurements of asteroid surface compositions) with analogous-wavelength laboratory measurements of OCs (EOCs and UOCs). We applied a radiative transfer model (Shkuratov et al. 1999) to the visible and near-IR spectra of both asteroid and meteorite samples following the same technique as in Brunetto et al. (2006), Vernazza et al. (2008, 2009, 2010) and Binzel et al. (2009). We used the two end-member minerals olivine (ol) and a low calcium pyroxene (low-Ca px), namely orthopyroxene, to determine their relative abundances and quantitatively evaluate compositions by the ratio ol/(ol + low-Ca px). It is worth noting that high-calcium pyroxenes such as diopside and augite do not show up in our asteroid spectra in larger abundances than seen in OCs. If this were the case, the two micron band of our asteroid and meteorite spectra would resemble that of Vesta and eucrites, which is not the case. Thus there is no evidence for partial melting on S-type surfaces at odds with the conclusions of Sunshine et al. (2004), Hardersen et al. (2006), and Gietzen et al. (2012).

We first applied this model to the ordinary chondrite meteorite spectra, namely we determined the ol/(ol + low-Ca px) ranges as a function of petrologic type for each group of ordinary chondrites (H, L, LL). In the case of UOCs, we applied the model to the near-IR wavelength range only, as the visible part of the spectrum is, in most cases, strongly affected by terrestrial weathering (see Appendix Fig. 3). We previously showed (Vernazza et al. 2010) that the application of our model to the shorter near-IR range (0.8-2.5 micron instead of 0.4-2.5 micron) only increases the error in the ol/(ol+opx) ratio by 2%. Our model results are in agreement with measurements by independent techniques (Menzies et al. 2005, Dunn et al. 2010) (Mössbauer spectra, XRD). Our new spectral data (Fig. 1) indicate that the ol/(ol + low-Ca px) ratio is the same for type 3.0-3.4 OCs [ol/(ol + low-Ca px) >65] with overall spectral properties that are surprisingly similar to those of equilibrated LL chondrites (see Appendix Figure A3). The spectral analysis further reveals that the spectral properties become distinct from those for low grade UOC for H chondrites and, to a lesser extent, L chondrites with increasing metamorphism, while they remain similar to those for low grade UOC for LL chondrites (see A3 and Appendix Figure A4 for a possible explanation of this trend). As a result, assuming an initial onion shell structure for H-like bodies, one could use spectroscopy to distinguish the outer shell from the metamorphosed interior (since type 3.0-3.4 H chondrites have ol/(ol + low-Ca px) ratios>65 and type 3.6-6 H chondrites have ol/(ol + low-Ca px) ratios<65).

We applied the same radiative transfer model to our asteroid population. To account for spectral reddening (if present) due to space weathering processes, we used a space weathering model (Brunetto et al. 2006). Figure 2 shows the distribution of inferred ol/(ol + low-Ca px) values for OCs and our main-belt sample. Surprisingly, the distribution for our asteroid sample appears to be bimodal (see A4 and Appendix Fig. A5), and not continuous or unimodal. The dip test (Hartigan & Hartigan 1985) - which measures multimodality in a sample by the maximum difference, over all sample points, between the empirical distribution function, and the unimodal distribution function that minimizes that maximum difference - gives a 99% confidence level against unimodality. This is not the first time that two spectrally distinct groups (i.e. a bimodality) are observed for relatively similar compositions among solar system small bodies. Emery et al. (2011) reported the presence of two compositional groups among the Jupiter Trojans while Tegler & Romanishin (1998) and later on Peixinho et al. (2012) reported the presence of two color groups among small TNOs and Centaurs.

We further notice that the bimodality of asteroid compositions is seen at all sizes, including large ones (D>100km), implying that it has a primordial origin (Appendix Fig. A6). Finally, we notice a difference in the size distribution between the two peaks (bodies in the left peak are on average smaller – see Appendix Fig. A6 and Appendix Fig. A7), which suggests, independently of the compositions, a different origin between the two peaks.

Asteroid-meteorite connections are inferred from the spectral bimodality. We find that in both their intrinsic spectral properties (Fig. 3) and in our mineralogical analysis (Fig. 2), asteroids in the left peak correspond to the parent bodies of H chondrites *with equilibrated surfaces* and asteroids in the right peak to the parent bodies of LL chondrites *with either equilibrated or unequilibrated surfaces*, and possibly also the parent bodies of H and L chondrites *with unequilibrated surfaces* (we will show below that the latter case is unlikely).

**4) Implications**

In this section, we present several proposed new constraints regarding the formation and evolution of ordinary chondrite parent bodies that are directly derived under the assumption that our compositional analysis gives correct inferences for asteroid-meteorite connections.

**4.1 Asteroid "clones" as a natural outcome of planetesimal formation**

Our broadened spectral survey indicates the presence of two compositional groups (shown by the bimodality) among S-type asteroids, namely (i) several objects have the same spectral signature (and quantitatively the same surface composition) as asteroid (6) Hebe (Fig. 4) and *equilibrated* H chondrites and (ii) several others have Flora-like (*8 Flora*) and LL chondrite-type spectral signatures. Collisions cannot be responsible for the existence of these two groups as these compositionally similar asteroids are found at large sizes (D>100km) and are unlikely to have experienced disruptive collisions since the formation of the asteroid belt (Bottke et al. 2005a, 2005b). Our observations thus indicate that "clones", namely identical compositions among multiple large asteroids (D≈100-200 km), are a natural outcome of planetesimal formation.

Lyra & Kuchner 2013 recently showed that due to interactions between gas and dust, a disk might, under the right conditions, produce narrow rings on its own without planets being needed. Ring-like structures, which are thus proposed as a natural step in the evolution of a protoplanetary disk, would naturally lead to the formation of several compositionally similar asteroids. The observation of two compositional groups only among S-type asteroids - implying that prior to accretion the disk must be locally quite homogeneous - appears very coherent with those findings. Our results are also consistent with current models of planetesimal formation via secular gravitational instabilities with turbulent stirring, which predict the formation of clans of chemically homogeneous planetesimals (e.g., Youdin 2011).

Our observations of several compositionally similar asteroids do open the possibility that a given meteorite group (e.g. H chondrites) can have more than one parent body, but do not imply that this is necessarily the case. Indeed, we want to stress here that it is not incompatible with our results that meteorite measurements should imply that a large fraction of meteorites in each OC group come from the same parent body. About 50% of the H chondrites with measured cosmic ray exposure ages have ages between 7 and 8 Myrs suggesting a similar origin (Eugster et al. 2006). Similarly, about two thirds of L chondrite meteorites were heavily-shocked and degassed with $^{39}$Ar–$^{40}$Ar ages near 470 Myr (Korochantseva et al., 2007) implying a common parent body for those meteorites (Haack et al. 1996).

It is worth noting that none of the properties that define the OC groups actually requires a single parent-body, not even oxygen isotopic signatures, which have been shown to be directly related to petrography (Zanda et al. 2006), more specifically to the relative proportions of reduced and oxidized chondrules. Chondrules constitute 80% of OC materials and therefore control their chemistry and mineralogy, as well as their oxygen isotopic signatures (Zanda et al. 2006), hence both the OC group to which they belong and their spectral properties. If groups of asteroids with similar spectra are a product of accretion, namely if they accreted from the same reservoir of chondrules and matrix, it is thus expected that they will also share their oxygen isotopic signatures.

Finally, we want to stress that asteroid (6) Hebe has many twins - including families - that lie as close or even closer (Fig. 5) to a resonance (3:1 resonance in particular). This implies that there are several alternatives to asteroid Hebe as plausible source(s) for H chondrites.

### 4.2 Surfaces of S-type asteroids as exposed interiors

Our deduction of *equilibrated* surface compositions on current day left-peak S-type asteroids [ol/(ol + low-Ca px)<65; *as the left peak (Fig. 2 and Fig. 3) is analogous to thermally metamorphosed (type 3.6 to 6; minimum T>~400˚C) H and possibly L chondrites*] is unexpected since onion-skin models for ordinary chondrite parent bodies (e.g., Ghosh et al. 2003, Henke et al. (2012a,b; 2013)) predict original surfaces of *unequilibrated* ordinary chondrite material (type 3.0 to 3.4). Equilibrated ordinary chondrite material is formed in volumetrically large abundance, but in the context of the onion skin models resides entirely in the metamorphosed interior of the original asteroid. Thus we propose the left peak is composed of asteroids whose surfaces are not primordial; rather their ol/(ol + low-Ca px) ratios reveal that today we are seeing the exposed interiors.

While many D<~100km sized bodies in the left peak are likely collisional fragments (Bottke et al. 2005a, Consolmagno et al. 2008, Morbidelli et al. 2009, Carry et al. 2012 and references therein) – thus naturally explaining that their surface composition is best matched by previously interior thermally metamorphosed chondrites (type 3.6 to 6), large bodies (D>100km) are more likely to be intact survivors capable of preserving at least some of their original surface compositions. Indeed, both current collisional models (Bottke et al. 2005a, Morbidelli et al. 2009) as well as the recent measurements performed by Rosetta on the asteroid Lutetia (Pätzold et al. 2011, Sierks et al. 2011) and by Dawn on the asteroid Vesta (Marchi et al. 2012) minimize the possibility that large bodies (D>100km) are collisional fragments themselves; instead these objects are predominantly primordial (undisrupted) planetesimals, which means they are unlikely to have undergone a catastrophic disruption since their formation. Specifically, an impactor population capable of producing widespread collisional disruption among D = 100-200 km asteroids would:

a) strongly affect asteroids like Vesta by creating ~9-10 basin-forming events equivalent to the one that made the D ~ 500 km Rheasilvia basin. Right now, we know of 2 such basins on Vesta (Marchi et al. 2012). Each basin is also associated with a band of deep equatorial troughs as measured from the central axis passing through the center of the basin. For now, it seems highly unlikely that one could (i) hide 7 additional such basins on Vesta's surface and, at the same time (ii) avoid bringing lots of the diogenites to the surface. Note that relaxation of Vesta's surface may have occurred during the first ~100 Myrs of the solar system implying that primary basins may be no longer observable (Fu et al. 2013). As such, Vesta's actual surface may thus not be fully incompatible with an early period of more intense bombardment.

b) create enumerable asteroid families filled with large fragments. (A family is produced when a large asteroid undergoes a catastrophic collision, leaving behind numerous fragments with similar proper orbital elements.) Yet, the telltale evidence that should have been produced by such disruption events: numerous asteroid families with large fragments, is not obvious in the main belt today as we only see a few large S-type asteroid families, the oldest being ~2 Gyrs old (e.g. Koronis family). Thus, some mechanism would be needed to eliminate this evidence.

A natural explanation for the surfaces of D>100km left-peak S-type asteroids consisting mostly of material formed in the interior, is that impacts by small asteroids over the course of solar system history (D<10km; sub-catastrophic collisions) and in particular those that occurred during the first ~20 Myr of solar system formation (Ciesla et al. 2013), brought up deep material to the surface. Indeed, Ciesla et al. (2013) recently showed that early impacts on "warm" bodies (*up to 20 Myrs after the bodies were formed*) would be far more efficient then later ones on "cold" bodies (*500 Myrs after solar system formation or even later*) in bringing deep material to the surface. Following their work, early impacts during the first ~20 Myr of solar system formation – where the interior of OC parent bodies would still be at high temperature from the decay of short-lived nuclides such $^{26}$Al - would result in hot material from the deep interior (material mainly from 10-30 km depth in a D=200km body) being brought to, and flowing out over, the surface of the planetesimal, covering most of its surface. This contrasts with the effect of later impacts into cold planetesimals, with material from no deeper than ~10km below the surface (in a D=200km body) being slightly heated by the impact and then exposed around the point of impact.

For asteroids at the right of the left peak (ol/(ol + low-Ca px)>65), the ol/(ol + low-Ca px) ratios do not distinguish between a surface and interior origin. However, since the sizes of these bodies do not differ significantly from those in the left peak, [although they do comprise more large (D>75km) bodies - see Appendix Fig. A6, A7], a different collisional history seems unlikely. We therefore suggest that the surfaces of these bodies are also exposed interiors which would imply that bodies in the left peak are H-like bodies while bodies in the right peak are LL-like bodies only (the presence of H-like bodies with unequilibrated surfaces in the right peak thus seems unlikely). Compositionally equilibrated L chondrites are intermediate between H and LL. Thus the compositional gap (Fig. 2) between the two peaks (60< ol/(ol + low-Ca px)<72) implies that asteroids with L chondrite surfaces must be rare.

In summary, our observations suggest that current day S-type asteroid surfaces are exposed interiors. Independent support to this finding is given by the metallographic cooling rates of H chondrites (discussed in section c) of the introduction). Finally, we would like to stress that S-type asteroids are not the only objects whose surfaces are exposed interiors. Indeed, this is also the case for the asteroid Vesta whose surface composition is best matched by howardites (Hiroi et al. 1994, De Sanctis et al. 2012), which are, by definition, a mixture of the primordial crust (eucrites) and of the underlying layer (diogenites).

**4.3 On the origin of biases in meteorite samples on Earth**

It is well known that atmospheric entry biases our meteorite collections by preferentially selecting the denser and more compact objects, which explains for example why ordinary chondrites are largely overrepresented in our collections with respect to carbonaceous chondrites such CI and CM chondrites. However, it still is unclear whether, at similar densities as it is the case for H, L and LL chondrites, these collections are fairly representative of the compositional diversity of the asteroid belt. As we have accumulated data for the majority of the most massive S-types (i.e., we sample more than 85% of the main belt mass for S-types) we can use our survey to make progress on this question.

Considering the fall statistics of the H (~34%), L (~37%) and LL (~8%) classes, both the paucity of L-like bodies among the asteroids and the dominance of LL-like bodies in the main belt population are very surprising. This implies that even at similar densities and tensile strength (which is typically the case for the H, L and LL chondrites) meteorite falls are not representative of the compositional diversity of the asteroid belt. A direct implication is that atmospheric entry is not the only factor that governs (i.e. biases) the inventory of our collections. Other factors such as the location of the source region versus the location of the main resonances, the nature of the source region (e.g., *asteroid collisional family versus cratering event on a large asteroid, size frequency distribution of the collisional family*) or the age of the source region (e.g., *age of the collisional family*) may play a fundamental role in the current statistics of meteorite falls.

It is clear that large bodies do not necessarily dominate the delivery of material to the Earth. For example, the largest main-belt asteroid Ceres (D~950km, ~25% of the mass of the main belt) may be unsampled by our meteorite collections as suggested by spectroscopy. Similarly, there are few lunar and Martian meteorites. Instead, asteroid families that are the result of catastrophic collisions producing an abundance of small fragments may be the main source of meteorites. This is possibly the case for the Gefion family that has been proposed (Nesvorny et al. 2009), based on dynamical arguments, to be the source of L chondrites. Here, we confirm that Gefion is the only known family whose composition is compatible with those of L chondrites, consequently reinforcing families as the most plausible sources of meteorites. The age of the families may also be a factor with older families possibly contributing less than younger ones to the meteorite flux as their smaller members may have been removed since a long time via the Yarkovsky effect (Rubincam 1995; Bottke et al. 2006). Future dynamical work will shed light on this question and as a byproduct may also explain the compositional difference between meteorites and NEAs (Vernazza et al. 2008; Deleon et al. 2010, Dunn et al. 2013).

**4.4 Fast accretion for the H parent bodies**

The accretion duration for asteroids, from the disk phase to their final size, is a fundamental and still open question. Both the size distribution of main belt asteroids and the most recent planetesimal formation models predict that asteroids formed big, namely that the size of solids in the proto-planetary disk ''jumped'' from sub-meter scale to multi-kilometer scale, without passing through intermediate values (Johansen et al. 2007, Cuzzi et al. 2008, Morbidelli et al. 2009). Here, we show that the surface composition of the members of H-like families is consistent with those predictions, thus providing another piece of evidence for the fast accretion scenario of asteroids. However, it should be noted that the present result does not allow distinguishing between instantaneous accretion and a growth over $\sim 10^5$ years and that it is highly dependent on current thermal models (e.g., Henke et al. 2012a,b; Henke et al. 2013). Specifically, it relies on the assumption that the main source of asteroid heating (metamorphism) was the decay of short-lived radioactive nuclides leading to an onion-shell structure for early asteroid parent bodies. If future findings demonstrate that this assumption is wrong, the present result will need to be re-interpreted.

As shown by Ghosh et al. (2003), the internal structure and therefore the thickness of the primordial unheated crust of type 3.0-3.4 material is directly related to the duration of accretion (Ghosh et al. 2003, Henke et al. 2012a,b; Henke et al. 2013):

- in the case of rapid accretion (instantaneous accretion or growth over $\sim 10^5$ years), asteroids have been quasi metamorphosed throughout (onion shell structure) and formed with a thin crust of type 3.0-3.4 material (<~5km assuming a D=200km sized parent body; see Ciesla et al. 2013 and references therein) (Fig. 6);

- in the case of a long duration of accretion (incremental accretion over at least 1Myr), only the cores have been metamorphosed leaving primordial bodies with a thick outer shell of type 3.0-3.4 material (>~20km assuming a D=200km sized parent body, more than ~40% in volume) (Fig. 6).

This difference of the internal structure as a function of duration of accretion is particularly interesting in the case of H-like bodies as one can use spectroscopy to distinguish the outer shell from the metamorphosed interior and thus to estimate the primordial percentage (in volume percent) of type 3.0.3-4 material in H-like bodies. To estimate this percentage, asteroid families are an ideal laboratory, as they allow us to access the interior composition of primordial parent bodies.

In the case of H-like asteroid families, we would expect to observe a clear compositional variation among the fragments in the case of a long duration of accretion as the volume of type 3.0-3.4 material (~40%) would be roughly equivalent to the volume of type 3.6-6 material (~60%). Specifically, we would expect to observe an extended compositional range, namely some members with a type 3.0-3.4 composition (LL-like), some other members with a type 3.6-6 composition (equilibrated H-like) and possibly some members being mixtures of type 3.0-3.4 and 3.6-6 materials (L-like, see Fig. 6). This is however not the case as families in the H peak (Agnia, Merxia, and Koronis) show little compositional variation among their members; in addition all the members of each family have a composition that is compatible with equilibrated H chondrites only. We estimate that type 3.0-3.4 materials can only represent a small fraction of the current surface material (<~15%), as for more than ~15%, there would be an obvious shift between the asteroid compositions and the compositions of equilibrated chondrites. Our observations are therefore consistent with an initially thin crust of type 3.0-3.4 materials (<~5km assuming a D=200km sized parent body) and thus with a short duration of accretion (<0.3 Myr; see Ghosh et al. 2003).

The small fraction of type 3.0-3.4 material implied by our observations is consistent with meteorite fall statistics (type 3 OCs represent ~15% of the falls and type 3.0-3.4 OCs less than 5%; see Hutchison 2004) while the short duration of accretion implied by our observations is consistent with current models of planet formation through streaming and gravitational instabilities (Youdin & Goodman 2005, Johansen et al. 2007, Chiang & Youdin 2010) that show that bodies of several hundred kilometers in size form on the time-scale of a few orbits (Johansen et al. 2011, Youdin 2011, Johansen et al. 2012).

**4.5 Size sorting in the disk as the origin of the H and LL formation locations?**

The distribution of asteroids across the main belt has been studied for decades to understand the current compositional distribution and what that tells us about the formation and evolution of our Solar System (Gradie & Tedesco 1982, DeMeo and Carry 2013, 2014). The result of those studies has shown the existence of a global compositional heliocentric gradient in the asteroid belt (S-types in the inner part; C-types in the outer part; Gradie & Tedesco 1982, DeMeo & Carry 2013, 2014). However, the compositional gradient among S-types, that is the respective formation locations of the H, L and LL parent bodies remain unknown.

We computed the average semi-major axis of asteroids in the two peaks in order to trace any initial difference in terms of formation location between the H and LL parent bodies. Since small objects (D<20km) are sensitive to the Yarkovsky effect, which affects their semi-major axis, the current semi major axis of D<20km planetesimals is quite different from their initial one; on the contrary, larger objects have not been moved from their current location since the last episode of major migration (e.g., Nice and Grand Tack models; Gomes et al. 2005, Tsiganis et al. 2005, Morbidelli et al. 2005, Walsh et al. 2011). We therefore restricted our sample to asteroids with D>30km (to be far enough from the D=20km limit) in the two peaks.

We found that LL-like bodies are located, on average, closer to the Sun (2.56 ± 0.04 AU) than H-like bodies (2.66 ± 0.03 AU) (Fig. 7). Note that as in the case of other compositional classes (e.g., S- and C-types), we observe a strong overlap in heliocentric distance between these two groups, which suggests that radial mixing of planetesimals after their formation must have been significant, a likely consequence of giant planet migrations (e.g., Walsh et al. 2011). Since migration models (e.g., Walsh et al. 2011) for small bodies and planets that try to reproduce the architecture of the solar system conserve the initial relative positions of the bodies with respect to the Sun, LL-like bodies are likely to have formed closer to the Sun than H-like ones.

Since chondrules in OCs are size-sorted, with LL chondrites containing on average larger chondrules than H chondrites (Kuebler et al. 1999, Zanda et al. 2006), we explore here whether size sorting of chondrules in the young protoplanetary disk might be an efficient physical mechanism that would naturally explain this observed difference in formation locations.

In a gaseous disk (Adachi et al. 1976; Weidenshilling 1977), solids with different sizes (and densities) have a different aerodynamic coupling with the gas, which automatically implies that they acquire non-zero relative radial velocities. In a Minimum Mass Solar Nebula model for the early Solar Nebula (Hayashi 1981), a chondrule with mean size (D=0.6mm) as seen in LL chondrites is expected to drift toward the Sun a few cm/s faster than a chondrule with mean size (D=0.3mm) as in H chondrites. Therefore, if one assumes that the H and LL parent bodies formed at a similar epoch, the slight shift in their formation location favors the idea of an initial turbulent concentration (Cuzzi et al. 2008) in the protoplanetary disk that produced dense zones of aerodynamically size-sorted particles. In such a case, larger chondrules would have been pushed inwards with respect to smaller ones, exactly as predicted by current models and in agreement with recent observations of protoplanetary disks (Perez et al. 2012). Interestingly, metal grains in OCs are located in between chondrules (rather than within as in CCs; see Fig A1 in appendix). Their aerodynamic properties are closer to those of the smaller chondrules (Kuebler et al. 1999), which may at the same time provide a simple explanation of the relative metal enrichment of H-chondrites. The chemistry/composition of a given OC class with respect to the other two OC classes may thus have more to do with the mean size of its chondrules (and possibly metal grains) than its formation location as suggested by Zanda et al. (2006). Our results may thus highlight the importance of transport processes in the disk prior to accretion on the *local* compositional heliocentric gradient in the asteroid belt.

**5 Conclusions and unresolved issues regarding the formation of OC parent bodies**

We conducted an extensive spectroscopic survey of 83 main belt S-type asteroids and 3 new S-type families, obtaining the biggest spectral data set yet assembled for the largest S-type asteroids (95% of all objects larger than 60km). In parallel, we built up the existing database of ordinary chondrite laboratory spectral measurements, which now spans a much broader range of temperature history (from unheated to significantly metamorphosed) than previously analyzed. This allowed us to discover several unexpected and fundamental results. First, most S-type asteroids, including large ones (D≈100-200 km), are distributed into two well-defined compositional groups, Hebe-like and Flora-like (H-like and LL-like). This indicates that identical compositions among multiple asteroids are a natural outcome of planetesimal formation and makes it possible that meteorites within a given class originate from multiple parent bodies. Second, the surfaces of nearly all of these asteroids (up to 200 km) show the same compositional characteristics as high temperature meteorites that were metamorphosed in the interiors of planetesimals. For such interior fragment compositions to be exposed on asteroid surfaces today, it is necessary that asteroids were thermally heated throughout, in which case their formation process must have been rapid. Last, we find that radial mixing of disk material prior to accretion, the size-sorting of chondrules in particular, has had consequential effects on the local heliocentric compositional gradient as shown by the LL chondrite parent bodies having formed closer to the Sun than those of H chondrite.

Future work may (i) explain the correlation between an object's size and composition that is observed in the H peak (see A.5 and Appendix Fig. 8), (ii) determine the internal structure (in terms of petrologic types) of ordinary chondrite parent bodies as a function of the parent body's size (Hebe being not necessarily the parent body of H chondrites, a D~100km sized parent body should be envisioned along a D~200km sized one) and time of formation, (iii) explain the diversity of the fall statistics between H, L and LL chondrites as a function of petrologic type (62% of H chondrites are type 4 and 5 while type 6 represent only 21%, 68% of Ls are type 5 and 6 and 59% of LLs are type 5 and 6; Hutchison 2004), and (iv) explore a possible link between chondrule size and/or the formation location, and the terminal size of a parent body in order to explain the smaller size of the H parent bodies [see A.6].

**References:**

Adachi, I., Hayashi, C., Nakazawa, K. The gas drag effect on the elliptical motion of a solid body in the primordial solar nebula. Prog. Theor. Phys. vol. 56, 1756-1771 (1976).

Akridge G., Benoit P. H. and Sears D. W. G. Regolith and megaregolith formation of H-chondrites: thermal constraints on the parent body. Icarus 132, 185–195 (1998).

Baker J., Bizzarro M. and Wittig N. Early planetisimal melting from an age of 4.5662 Gyr for differentiated meteorites. Nature 436, 1127–1131 (2005).

Bennett M. E. III and McSween H. Y., Jr. Revised model calculations for the thermal histories of ordinary chondrite parent bodies. Meteoritics & Planetary Science 31, 783–792 (1996).

Binzel, R. P. et al. Spectral properties and composition of potentially hazardous asteroid (99942) Apophis. Icarus 200, 480–485 (2009).

Bizzarro M., Baker J. A., Haack H., and Lundgaard K. L. Rapid time scales for accretion and melting of differentiated planetesimals inferred from $^{26}$Al-$^{26}$Mg chronometry. *The Astrophysical Journal* **632**, L41–44 (2005).

Bottke, W. F., Durda, D. D., Nesvorný, D., Jedicke, R., Morbidelli, A., Vokrouhlický, D., Levison, H. The fossilized size distribution of the main asteroid belt. Icarus 175, 111-140 (2005a).

Bottke, W. F., Durda, D. D., Nesvorný, D., Jedicke, R., Morbidelli, A., Vokrouhlický, D., Levison, H. F. Linking the collisional history of the main asteroid belt to its dynamical excitation and depletion. Icarus 179, 63-94 (2005b).

Bottke, W. F., Vokrouhlický, D., Rubincam, D. P., Nesvorný, D. The Yarkovsky and Yorp Effects: Implications for Asteroid Dynamics. Annual Review of Earth and Planetary Sciences 34, 157-191 (2006).

Brunetto, R. et al. Modeling asteroid surfaces from observations and irradiation experiments: The case of 832 Karin. Icarus 184, 327–337 (2006).

Bus, S. J. Compositional Structure in the Asteroid Belt: Results of a Spectroscopic Survey. Thesis, Massachusetts Inst. Technol. (1999).

Carry, B. Density of asteroids. Planetary and Space Science 73, 98-118 (2012).

Chiang, E. & Youdin, A. N. Forming Planetesimals in Solar and Extrasolar Nebulae. Annual Review of Earth and Planetary Sciences 38, 493-522 (2010).

Ciesla, F. J., Davison, T. M., Collins, G. S., O'Brien, D. P. Thermal consequences of impacts in the early solar system. Meteoritics & Planetary Science 48, 2559-2576 (2013).

Connolly Jr., H.C., Desch, S.J. On the origin of the ''kleine Kugelchen'' called chondrules. Chemie der Erde/Geochemistry 64, 95–125 (2004).

Consolmagno, G., Britt, D., Macke, R. The significance of meteorite density and porosity. Chemie der Erde – Geochemistry 68, 1-29 (2008).

Cuzzi, J. N., Hogan, R. C., Paque, J. M., Dobrovolskis, A. R. Size-selective Concentration of Chondrules and Other Small Particles in Protoplanetary Nebula Turbulence. The Astrophysical Journal 546, 496-508 (2001).

Cuzzi, J. N. & Weidenschilling, S. J. Particle-Gas Dynamics and Primary Accretion. In Meteorites and the Early Solar System II, D. S. Lauretta and H. Y. McSween Jr. (eds.), University of Arizona Press, Tucson, 353-381 (2006).

Cuzzi, J. N., Hogan, R. C., Shariff, K. Toward Planetesimals: Dense Chondrule Clumps in the Protoplanetary Nebula. The Astrophysical Journal 687, 1432-1447 (2008).

Davison, T. M., Ciesla, F. J., Collins, G. S. Post-impact thermal evolution of porous planetesimals. Geochimica et Cosmochimica Acta 95, 252-269 (2012).

De León, J., Licandro, J., Serra-Ricart, M., Pinilla-Alonso, N., Campins, H. Observations, compositional, and physical characterization of near-Earth and Mars-crosser asteroids from a spectroscopic survey. Astronomy and Astrophysics 517, id.A23, 25 pp. (2010).

DeMeo, F.E., Binzel, R.P., Slivan, S., Bus, S.J. An extension of the Bus asteroid taxonomy into the near-infrared. Icarus 202, 160–180 (2009).

DeMeo, F. E., Carry, B. The taxonomic distribution of asteroids from multi-filter all-sky photometric surveys. Icarus 226, 723-741 (2013).

DeMeo, F. E., Carry, B. Solar System evolution from compositional mapping of the asteroid belt. Nature 505, 629-634 (2014).

De Sanctis, M. C. et al. Spectroscopic Characterization of Mineralogy and Its Diversity Across Vesta. Science 336, 697- (2012).

Dodd, R. T., Jr., Koffman, D. M., van Schmus, W. R. A survey of the unequilibrated ordinary chondrites. *Geochimica et Cosmochimica Acta* **31**, 921-934 (1967).

Dunn, T. L., Cressey, G., McSween, H. Y., McCoy, T. J. Analysis of ordinary chondrites using powder X-ray diffraction: 1. Modal mineral abundances. Meteoritics and Planetary Science 45, 123-134 (2010a).

Dunn, T. L., McSween, H. Y. _jr., Jr., McCoy, T. J., Cressey, G. Analysis of ordinary chondrites using powder X-ray diffraction: 2. Applications to ordinary chondrite parent-body processes. *Meteoritics and Planetary Science* **45**, 135-156 (2010b).

Dunn, T. L., Burbine, T. H., Bottke, W. F., Clark, J. P. Mineralogies and source regions of near-Earth asteroids. Icarus 222, 273-282 (2013).

Elkins-Tanton, L. T., Weiss, B. P.; Zuber, M. T. Chondrites as samples of differentiated planetesimals. Earth and Planetary Science Letters 305, 1-10 (2011).

Emery, J. P., Burr, D. M., Cruikshank, D. P. Near-infrared Spectroscopy of Trojan Asteroids: Evidence for Two Compositional Groups. The Astronomical Journal 141, article id. 25 (2011).

Eugster, O., Herzog, G. F., Marti, K., Caffee, M. W. Irradiation Records, Cosmic-Ray Exposure Ages, and Transfer Times of Meteorites. Meteorites and the Early Solar System II, D. S. Lauretta and H. Y. McSween Jr. (eds.), University of Arizona Press, Tucson, 829-851 (2006).

Fu, R. R., Hager, B. H., Ermakov, A. I., Zuber, M. T. Early Viscous Relaxation of Asteroid Vesta and Implications for Late Impact-Driven Despinning. 44th Lunar and Planetary Science Conference, Contribution No. 1719, p.2115 (2013).

Gaffey, M.J., Bell, J.F., Brown, R.H., Burbine, T.H., Piatek, J., Reed, K.L., Chaky, D.A. Mineralogic variations within the S-type asteroid class. Icarus 106, 573–602 (1993).

Gaffey, M. J., Gilbert, S. L. Asteroid 6 Hebe: The probable parent body of the H-Type ordinary chondrites and the IIE iron meteorites. Meteoritics & Planetary Science 33, 1281-1295 (1998).

Gastineau-Lyons, H. K.; McSween, H. Y., Jr.; Gaffey, M. J. A critical evaluation of oxidation versus reduction during metamorphism of L and LL group chondrites,and implications for asteroid spectroscopy. Meteoritics & Planetary Science, 37, no. 1, pp. 75-89 (2002).

Ghosh A., Weidenschilling S. J., and McSween H. Y. Jr. Importance of the accretion process in asteroidal thermal evolution: 6 Hebe as an example. Meteoritics & Planet. Sci. 38, 711–724 (2003).

Ghosh, A., Weidenschilling, S. J., McSween, H. Y., & Rubin, A. In Meteorites and the Early Solar System II, ed. D. S. Lauretta, & H. Y. McSween Jr. (Tucson: Univ. of Arizona Press), 555-566 (2006).

Gietzen, K. M., Lacy, C. H. S., Ostrowski, D. R., Sears, D. W. G. IRTF observations of S complex and other asteroids: Implications for surface compositions, the presence of clinopyroxenes, and their relationship to meteorites. Meteoritics & Planetary Science 47, 1789-1808 (2012).

Gomes, R., Levison, H.F., Tsiganis, K., Morbidelli, A., 2005. Origin of the cataclysmic Late Heavy Bombardment period of the terrestrial planets. Nature 435, 466–469.

Gooding, J.L. Mineralogical aspects of terrestrial weathering effects in chondrites from Allan Hills, Antarctica. Lunar Planet. Sci., 1105–1122 (1982).

Gradie, J., Tedesco, E. Compositional structure of the asteroid belt. Science 216, 1405-1407 (1982).

Graf, T. & Marti, K. Collisional records in LL-chondrites. Meteoritics 29, 643-648 (1994).

Graf, T. & Marti, K. Collisional history of H chondrites. Journal of Geophysical Research 100, 21247-21264 (1995).

Grimm R. E. and McSween H. Y., Jr. Heliocentric zoning of the planetesimal belt by aluminum-26 heating. Science 259, 653–655 (1993).

Grossman, J. N., Rubin, A. E., Nagahara, H., King, E. A. Properties of chondrules. Meteorites and the early solar system. Tucson, AZ, University of Arizona Press, 619-659 (1989).

Haack, H., Farinella, P., Scott, E.R.D., Keil, K. Meteoritic, asteroidal, and theoretical constraints on the 500 Ma disruption of the L chondrite parent body. Icarus 119, 182–191 (1996).

Hardersen, P. S., Gaffey, M. J., Cloutis, E. A., Abell, P. A.; Reddy, V. Near-infrared spectral observations and interpretations for S-asteroids 138 Tolosa, 306 Unitas, 346 Hermentaria, and 480 Hansa. Icarus 181, 94-106 (2006).

Hartigan, J. A. & Hartigan, P. M. The Dip Test of Unimodality. The Annals of Statistics 13, 70-84, (1985).

Harrison K. P. and Grimm R. E. Thermal constraints on the early history of the H-chondrite parent body reconsidered. Geochimica et Cosmochimica Acta 74, 5410–5423 (2010).

Hayashi, C. Structure of the Solar Nebula, Growth and Decay of Magnetic Fields and Effects of Magnetic and Turbulent Viscosities on the Nebula. Progress of Theoretical Physics Supplement 70, 35-53 (1981).

Henke, S., Gail, H.-P., Trieloff, M., Schwarz, W. H., Kleine, T. Thermal evolution and sintering of chondritic planetesimals. Astronomy & Astrophysics 537, id.A45 (2012a).

Henke, S., Gail, H.-P., Trieloff, M., Schwarz, W. H., Kleine, T. Thermal history modelling of the H chondrite parent body. Astronomy & Astrophysics 545, id.A135, (2012b).

Henke, S., Gail, H.-P., Trieloff, M., Schwarz, W. H. Thermal evolution model for the H chondrite asteroid-instantaneous formation versus protracted accretion. Icarus 226, 212-228 (2013).

Hevey P. J. and Sanders S. A model for planetesimal meltdown by $^{26}$Al and its implications for meteorite parent bodies. *Meteoritics & Planetary Science* **41**, 95–106 (2006).

Hiroi, T., Pieters, C. M., Takeda H. Grain size of the surface regolith of asteroid 4 Vesta estimated from its reflectance spectrum in comparison with HED meteorites. Meteoritics 29, 394-396 (1994).

Huss, G. R., Rubin, A. E., Grossman, J. N. Thermal Metamorphism in Chondrites. In Meteorites and the Early Solar System II, D. S. Lauretta and H. Y. McSween Jr. (eds.), University of Arizona Press, Tucson, 567-586 (2006).

Hutchison, R. Meteorites: A Petrologic, Chemical and Isotopic Synthesis (Cambridge Univ. Press, 2004).

Jacquet, E., Gounelle, M., Fromang, S. On the aerodynamic redistribution of chondrite components in protoplanetary disks. Icarus 220, 162-173 (2012).

Jarosewich, E. Chemical analyses of meteorites - A compilation of stony and iron meteorite analyses. *Meteoritics* **25**, 323-337 (1990).

Johansen, A., Oishi, J. S., Mac Low, M.-M., Klahr, H., Henning, T., Youdin, A. Rapid planetesimal formation in turbulent circumstellar disks. Nature 448, 1022-1025 (2007).

Johansen, A., Klahr, H., Henning, Th. High-resolution simulations of planetesimal formation in turbulent protoplanetary discs. Astronomy & Astrophysics 529, id.A62 (2011).

Johansen, A., Youdin, A. N., Lithwick, Y. Adding particle collisions to the formation of asteroids and Kuiper belt objects via streaming instabilities. Astronomy & Astrophysics 537, id.A125 (2012).

Keil, K., Fredriksson, K. The Iron, Magnesium, and Calcium Distribution in Coexisting Olivines and Rhombic Pyroxenes of Chondrites. *Journal of Geophysical Research* **69**, 3487-3515 (1964).

Keil K., Stöffler D., Love S. G., and Scott E. R. D. Constraints on the role of impact heating and melting in planetesimals. Meteoritics & Planetary Science 32, 349–363 (1997).

Kelley, M. S., Wooden, D. H. The composition of dust in Jupiter-family comets inferred from infrared spectroscopy. Planetary and Space Science 57, 1133-1145 (2009).

Korochantseva, E.V., Trieloff, M., Lorenz, C.A., Buykin, A.I., Ivanova, M.A., Schwarz, W.H., Hopp, J., Jessberger, E.K. L-chondrite asteroid breakup tied to Ordovician meteorite shower by multiple isochron 40Ar–39Ar dating. Meteorit. Planet. Sci. 42, 113–130 (2007).
33

Krot, A. N. et al. Progressive alteration in CV3 chondrites: More evidence for asteroidal altération. Meteoritics & Planetary Science 33, 1065-1085 (1998).

Krot, T. V., Goldstein, J. I., Scott, E. R. D., and S. Wakita. Thermal Histories of H3–6 Chondrites and Their Parent Asteroid from Metallographic Cooling Rates and Cloudy · Taenite Dimensions. M&PSA 5372 (2012).

Kuebler, K. E. et al. Sizes and Masses of Chondrules and Metal-Troilite Grains in Ordinary Chondrites: Possible Implications for Nebular Sorting. Icarus 141, 96-106 (1999).

Love, S. G., Ahrens, T. J. Catastrophic Impacts on Gravity Dominated Asteroids. Icarus 124, 141-155 (1996).

Lyra, W. & Kuchner, M. Formation of sharp eccentric rings in debris disks with gas but without planets. Nature 499, 184-187 (2013).

Marchi. S, et al. The Violent Collisional History of Asteroid 4 Vesta. Science 336, 690- (2012).

Marsh C. A., Della-Giustina D. N., Giacalone J., and Lauretta D. S. Experimental tests of the induction heating hypothesis for planetesimals (abstract #2078). 37th Lunar and Planetary Science Conference. CD-ROM (2006).

Marti, K. & Graf, T. Cosmic-ray exposure history of ordinary chondrites. Annual Review of Earth and Planetary Sciences 20, 221-243 (1992).

Masiero, J. R., Mainzer, A. K., Grav, T., Bauer, J. M. et al. Main Belt Asteroids with WISE/NEOWISE. I. Preliminary Albedos and Diameters. The Astrophysical Journal 741, article id. 68 (2011).

Mason, B. Olivine composition in chondrites. *Geochimica et Cosmochimica Acta* **27**, 1011-1023 (1963).

McSween, H. Y., Jr., Patchen, A. D. Pyroxene thermobarometry in LL-group chondrites and implications for parent body metamorphism. Meteoritics 24, 219-226 (1989).

McSween H. Y., Jr., Ghosh A., Grimm R. E., Wilson L., and Young E. D. Thermal evolution models of planetesimal. In Planetesimals III, edited by Bottke W. F. Jr., Cellino A., Paolicchi P. and Binzel R. P. Tucson, Arizona: The University of Arizona Press. pp. 559–571 (2002).

McSween H. Y., Jr. and Labotka, T. C. Oxidation during metamorphism of the ordinary chondrites, *Geochim. Cosmochim. Acta*, vol. 57, no. 5, p. 1105-1114. (1993)

Menzies, O. N., Bland, P. A., Berry, F. J., Cressey, G. A Mössbauer spectroscopy and X-ray diffraction study of ordinary chondrites: Quantification of modal mineralogy and implications for redox conditions during metamorphism. Meteoritics & Planetary Science 40, 1023–1042 (2005).

Merk R., Breuer D., and Spohn T. Numerical modeling of $^{26}$Al induced radioactive melting of planetesimals considering accretion. *Icarus* **159**, 183–191 (2002).

Miyamoto M., Fujii N., and Takeda H. Ordinary chondrites parent body: An internal heating model. Proceedings, 12th Lunar and Planetary Science conference. 1145–1152 (1981).

Miyamoto M. Thermal metamorphism of CI and CM carbonaceous chondrites: An internal heating model. Meteoritics 26, 111–115 (1991).

Monnereau, M., Toplis, M. J., Baratoux, D., Guignard, J. Thermal history of the H-chondrite parent body: Implications for metamorphic grade and accretionary time-scales. Geochimica et Cosmochimica Acta 119, 302-321 (2013).

Morbidelli, A., Levison, H.F., Tsiganis, K., Gomes, R., 2005. Chaotic capture of Jupiter's Trojan asteroids in the early Solar System. Nature 435, 462–465.

Morbidelli, A. et al. Asteroids were born big. Icarus 204, 558-573 (2009).

Mostefaoui S., Lugmair G. W., and Hoppe P. $^{60}$Fe: A heat source for planetary differentiation from a nearby supernova explosion. *The Astrophysical Journal* **625**, 271–277 (2005).

Nakamura et al. Itokawa Dust Particles: A Direct Link Between S-Type Asteroids and Ordinary Chondrites. Science 333, 1113- (2011).

Neumann W., Breuer D. and Spohn T. Differentiation and core formation in accreting planetesimals. Astron. Astrophys. 543, A141 (2012).

Nesvorný, D., Vokrouhlický, D., Morbidelli, A., Bottke, W. F. Asteroidal source of L chondrite meteorites. Icarus 200, 698-701 (2009).

Pätzold, M. et al. Asteroid 21 Lutetia: Low Mass, High Density. Science 334, 491-492 (2011).

Peixinho, N., Delsanti, A., Guilbert-Lepoutre, A., Gafeira, R., Lacerda, P. The bimodal colors of Centaurs and small Kuiper belt objects. Astronomy & Astrophysics 546, id.A86 (2012).

Perez, L. et al. Constraints on the Radial Variation of Grain Growth in the AS 209 Circumstellar Disk. The Astrophysical Journal Letters 760, article id. L17 (2012).

Pieters, C. M. et al. Space weathering on airless bodies: Resolving a mystery with lunar samples. Meteorit. Planet. Sci. 35, 1101–1107 (2000).

Rayner, J. T. et al. A medium-resolution 0.8–5.5micron spectrograph and imager for the NASA Infrared Telescope Facility. Publ. Astron. Soc. Pacif. 115, 362–382 (2003).

Rubin A. E. Petrologic evidence for collisional heating of chondritic asteroids. Icarus 113, 156–167 (1995).

Rubin, A. E. Mineralogy of meteorite groups. Meteoritics 32, 231-247 (1997).

Rubin A. E. Chromite-plagioclase assemblages as a new shock indicator; Implications for the shock and thermal histories of ordinary chondrites. Geochimica et Cosmochimica Acta 67, 2695–2709 (2003).

Rubin A. E. Postshock annealing and postannealing shock in equilibrated ordinary chondrites: Implications for the thermal and shock histories of chondritic asteroids 1. Geochimica et Cosmochimica Acta 68:673–689 (2004).

Rubin, A. E. Relationships among intrinsic properties of ordinary chondrites: Oxidation state, bulk chemistry, oxygen-isotopic composition, petrologic type, and chondrule size. Geochimica et Cosmochimica Acta 69, 4907-4918 (2005).

Rubincam, D. P. Asteroid orbit evolution due to thermal drag. J. Geophys. Res. 100, 1585–1594 (1995).

Sahijpal S., Soni P., and Gupta G. Numerical simulations of the differentiation of accreting planetesimals with $^{26}$Al and $^{60}$Fe as the heat sources. Meteoritics & Planetary Science 42, 1529–1548 (2007).

Schilling et al. Is Human Height Bimodal? *The American Statistician* **56**, 223-229 (2002).

Scott, E.R.D, Love, S.G. and Krot, A.N. Formation of chondrules and chondrites in the protoplanetary nebula. *In Chondrules and the Protoplanetary Disk*, (R. H. Hewins, R. H. Jones and E. R. D. Scott eds.), Cambridge University Press, 87-96 (1996).

Scott E. R. D., Krot T. V., Goldstein J. I. and Taylor G. J. Thermal and impact history of H chondrites: was the onionshell punctured by impacts during metamorphism? Meteoritics and Planetary Science Abstr. 5516. 74th Annual Meteoritical Society Meeting (2011).

Scott, E. R. D., Krot, T. V., Goldstein, J. I., Herzog, G. F. Impact and Thermal History of the H Chondrite Parent Body Inferred from Fe-Ni Metal and Ar/Ar Ages. Meteoritics and Planetary Science Abstr. 5346. 76th Annual Meteoritical Society Meeting (2013).

Scott, E.R.D. ; Krot, T.V. ; Goldstein, J.I. and Wakita, S. Thermal and impact history of the H chondrite parent asteroid during metamorphism: Constraints from metallic Fe–Ni. *Geochim. Cosmochim. Acta* 136, 13–37 (2014).

Sears, D. W., Grossman, J. N., Melcher, C. L., Ross, L. M., Mills, A. A. Measuring metamorphic history of unequilibrated ordinary chondrites. *Nature* **287**, 791-795 (1980).

Shkuratov, Y., Starukhina, L., Hoffmann, H. & Arnold, G. A model of spectral albedo of particulate surfaces: Implications for optical properties of the Moon. Icarus 137, 235–246 (1999).

Sierks, H., Lamy, P., et al. Images of Asteroid 21 Lutetia: A Remnant Planetesimal from the Early Solar System. Science 334, 487-490 (2011).

Sunshine, J. M., Bus, S. J.; McCoy, T. J., Burbine, T. H., Corrigan, C. M., Binzel, R. P. High-calcium pyroxene as an indicator of igneous differentiation in asteroids and meteorites. Meteoritics & Planetary Science 39, 1343-1357 (2004).

Tachibana S. and Huss G. R. The initial abundance of $^{60}$Fe in the solar system. *The Astrophysical Journal* **588**, L41–44 (2003).

Taylor, G. J., Maggiore, P., Scott, E. R. D., Rubin, A. E., Keil, K. Original structures, and fragmentation and reassembly histories of asteroids - Evidence from meteorites. Icarus 69, 1-13 (1987).

Tegler, S. C., Romanishin, W. Two distinct populations of Kuiper-belt objects. Nature 392, 49-51 (1998).

Trieloff M., Jessberger E. K., Herrwerth I., Hopp J., Fieni C., Ghelis M., Bourot-Denise M. and Pellas P. Structure and thermal history of the H-chondrite parent asteroid revealed by thermochronometry. Nature 422, 502–506 (2003).

Tsiganis, K., Gomes, R., Morbidelli, A., Levison, H.F., 2005. Origin of the orbital architecture of the giant planets of the Solar System. Nature 435, 459–461.

Urey H. C. The cosmic abundances of potassium, uranium, and thorium and the heat balances of the earth, the Moon, and Mars. Proceedings of the National Academy of Science 41, 127–144 (1955).

Van Schmus, W. R. & Wood, J. A. A chemical-petrologic classification for the chondritic meteorites. Geochim. Cosmochim. Acta 31, 747–765 (1967).

Vernazza, P. et al. Compositional differences between meteorites and near-Earth asteroids. Nature 454, 858–860 (2008).

Vernazza, P., Binzel, R. P., Rossi, A., Fulchignoni, M. & Birlan, M. Solar wind as the origin of rapid weathering of asteroid surfaces. Nature 458, 993–995 (2009).

Vernazza, P., Carry, B., Emery, J., Hora, J.L., Cruikshank, D., Binzel, R.P., Jackson, J., Helbert, J., Maturilli, A. Mid-infrared spectral variability for compositionally similar asteroids: Implications for asteroid particle size distributions. Icarus 207, 800–809 (2010).

Vernazza, P., Fulvio, D., Brunetto, R., Emery, J. P., Dukes, C. A., Cipriani, F., Witasse, O., Schaible, M. J., Zanda, B., Strazzulla, G., Baragiola, R. A. Paucity of Tagish Lake-like parent bodies in the Asteroid Belt and among Jupiter Trojans. Icarus 225, 517-525 (2013).

Walsh, K. J. et al. A low mass for Mars from Jupiter's early gas-driven migration. Nature 475, 206-209 (2011).

Weidenschilling, S. J. Aerodynamics of solid bodies in the solar nebula. Monthly Notices of the Royal Astronomical Society 180, 57-70 (1977).

Wood, J. A. Planetary science: Of asteroids and onions. Nature 422, 479-481 (2003).

Yomogida, K. & Matsui, T. Multiple parent bodies of ordinary chondrites. Earth and Planetary Science Letters 68, 34-42 (1984).

Youdin, A. N. & Goodman, J. Streaming Instabilities in Protoplanetary Disks. The Astrophysical Journal 620, 459-469 (2005).

Youdin, A. N. On the Formation of Planetesimals Via Secular Gravitational Instabilities with Turbulent Stirring. The Astrophysical Journal 731, article id. 99 (2011).

Zanda, B. Chondrules. *Earth and Planetary Science Letters* **224**, 1-17 (2004).

Zanda, B., Hewins, R. H.;,Bourot-Denise, M., Bland, P. A., Albarède, F. Formation of solar nebula reservoirs by mixing chondritic components. Earth and Planetary Science Letters 248, 650-660 (2006).

**Acknowledgements**

We thank the referee for his pertinent and constructive remarks. The research leading to these results has received funding from the European Community's Seventh Framework Programme. The work by the MIT co-authors was supported by the National Science Foundation. The near-infrared data were acquired by the authors operating as Visiting Astronomers at the Infrared Telescope Facility, which is operated by the University of Hawaii under Cooperative Agreement with the National Aeronautics and Space Administration, Science Mission Directorate, Planetary Astronomy Program. We warmly thank the Antarctic meteorite collection for providing the ordinary chondrite samples. We thank Cecilia Satterwhite for preparing the samples. We thank Zibulle for producing Figure 6's artwork. We thank Tom Burbine for useful discussions and encouragement. This paper is dedicated to the late Paul Pellas who energetically defended the idea that S-type asteroids are the parent bodies of OCs.

**Figures**

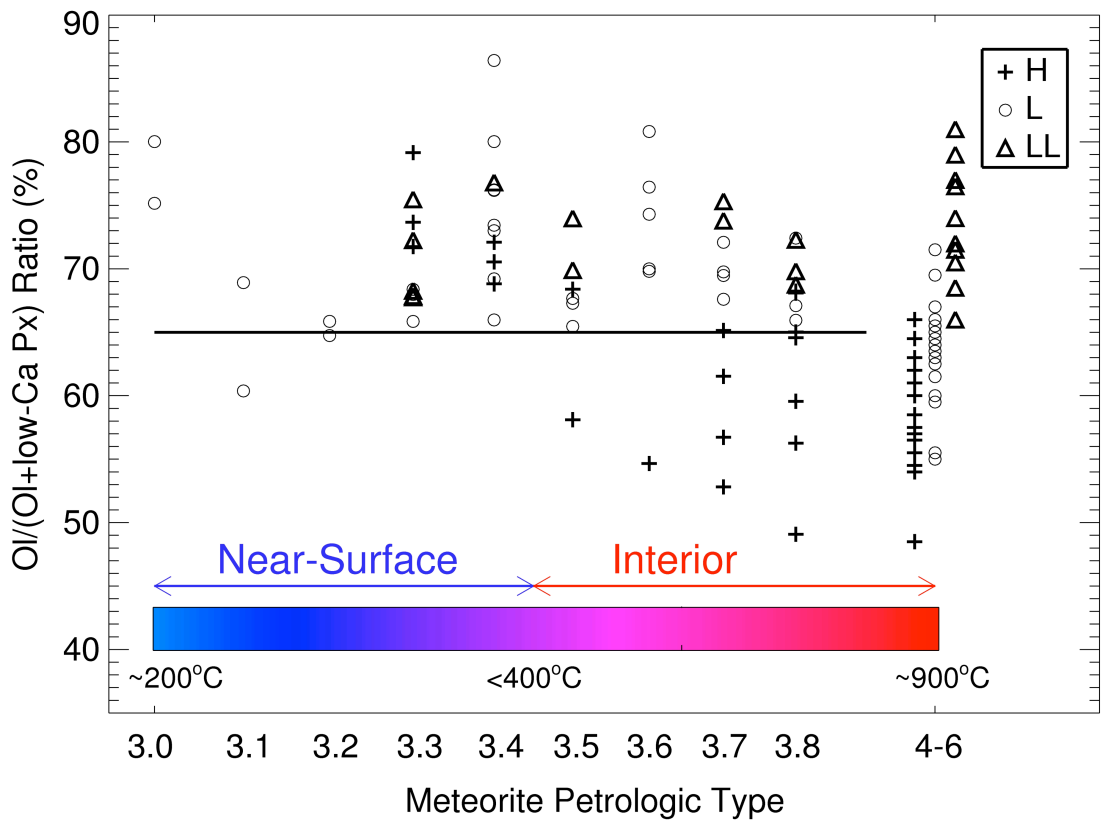

**Figure 1: Composition of ordinary chondrites as a function of petrologic type.** A planetary embryo that is internally heated (as in the case of asteroids, by the decay of short-lived nuclides) reaches higher maximum metamorphic temperatures (indicated on the plot) in its centre resulting in higher petrologic types; the outer layers, on the contrary, cool faster and reach lower metamorphic temperatures resulting in lower petrologic types. The spectra for 53 unequilibrated ordinary chondrites (UOCs; types 3.0 to 3.8) were measured at Brown University. These new results more than quintuple the number available UOC measurements, and unexpectedly show a uniform high olivine proportion for the lowest types (3.0-3.5). These least metamorphosed OCs are the best proxy we have of the primordial composition of the protoplanetary disk in this region of the Solar System. The spectra for equilibrated ordinary chondrites (EOCs; types 4-6) were retrieved from the RELAB database. (We display all EOCs on the same axis, as their breadth is homogenous across types 4-6 in terms of ol/(ol + low-Ca px) ratio; for clarity, the three classes are slightly shifted horizontally) UOCs for types 3.0 to 3.5 are olivine rich and their composition is relatively 'homogeneous' (same for H, L and LL). The same result was confirmed with 2 other approaches (Mössbauer spectra, XRD).

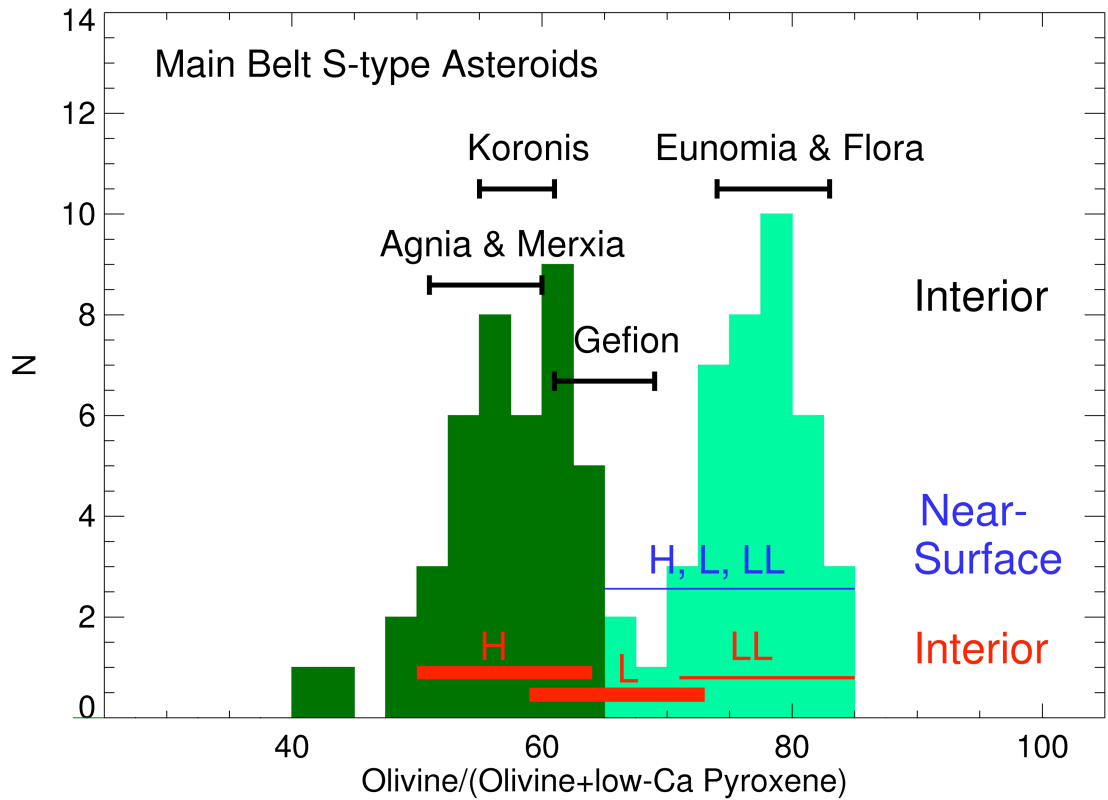

**Figure 2**: **Bimodal compositional distribution of main-belt S-type asteroids.** Our sample comprises 83 objects, including 54 out of the 56 main-belt S-types with D>60km. Objects belonging to collisional families are not included in the histogram counts. Instead, the compositional range for the 6 main asteroid families (Agnia, Merxia, Koronis, Gefion, Eunomia, Flora) are shown at the top. The compositional ranges for the individual ordinary chondrite classes of « Interior » samples (H, L and LL having petrologic types >3.5; temperature histories >400°C) are shown in red. The blue line denotes the compositional range for the least metamorphosed OCs (Types 3.0-3.5 ; temperatures <400°C) that are interpreted as surface samples. The thickness of the various compositional ranges for meteorites is proportional to their fall statistics. Finally, the diversity of S-type asteroids includes compositions outside the range of ordinary chondrites as seen for the two objects having ol/(ol + low-Ca px) < 45%.

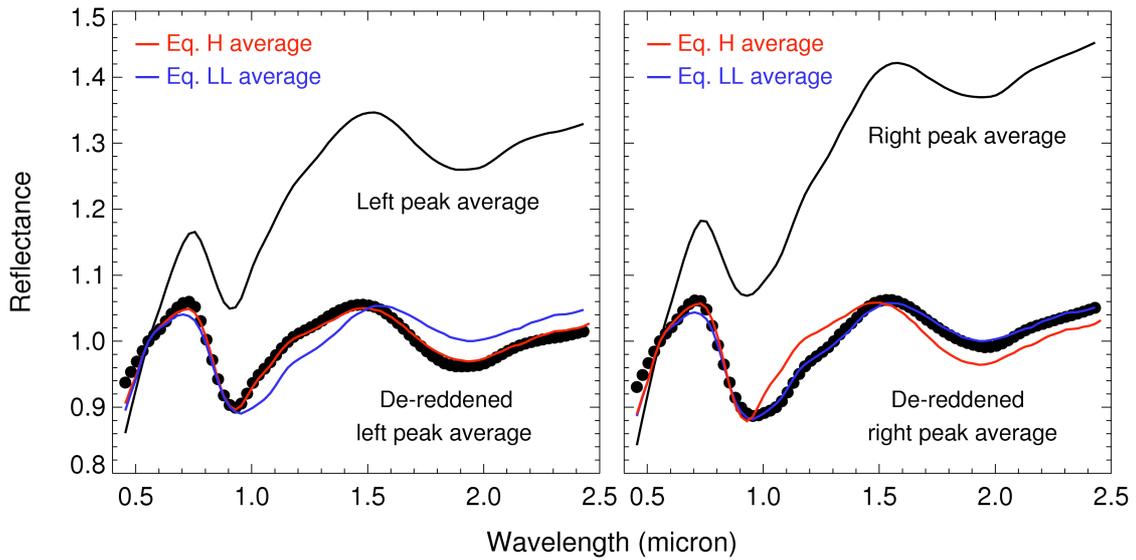

**Figure 3: Spectral comparisons of asteroids and meteorites.** Comparison between the visible to near-infrared spectral signatures of main belt S-type asteroids (left: asteroids in the left peak of the bimodality as revealed by our spectral analysis; right: asteroids in the right peak of the bimodality as revealed by our spectral analysis) and the average spectra of equilibrated (type 4 to 6) H and LL chondrite meteorites. Space weathering processes similar to those acting on the Moon (e.g., Pieters et al. 2000) redden and darken the ordinary condrite-like spectrum of a fresh asteroid surface, giving it the appearance of an S-type spectrum (e.g., Vernazza et al. 2008 and references therein). Thus, for a quantitative comparison between the asteroid and meteorite spectra, we de-reddened the average asteroid spectra with the space weathering model developped by Brunetto et al. (2006). In addition, we allowed the spectral contrast of the meteorite spectra to vary (as in Vernazza et al. 2013 for example). We find that asteroids in the left peak correspond to the parent bodies of H chondrites (*with equilibrated surfaces*) and asteroids in the right peak to the parent bodies of LL chondrites. Note that olivine-rich asteroids (right) are on average redder than "olivine-poor" ones, in agreement with previous findings (Vernazza et al. 2009).

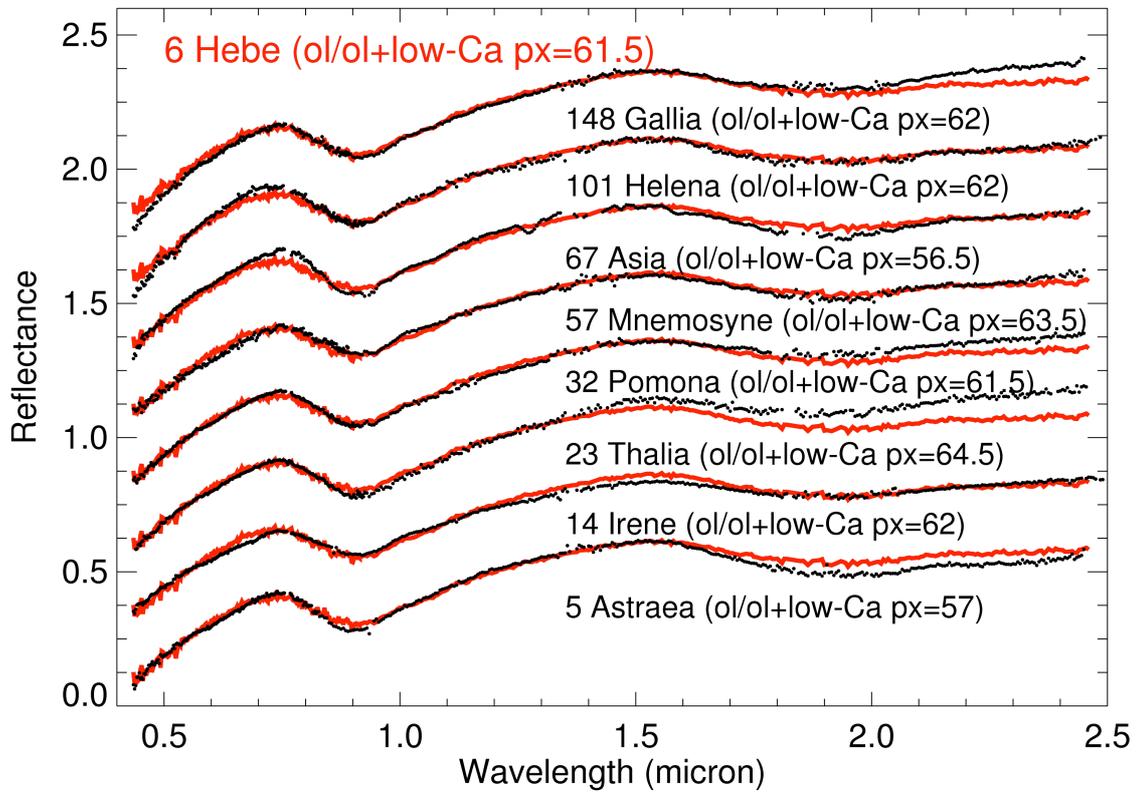

**Figure 4: Multiple parent bodies for a given meteorite class.** While asteroid 6 Hebe has been proposed as a single body source for H chondrites (Gaffey & Gilbert 1998), expanded spectral surveys now reveal multiple different asteroids as viable H chondrite candidates in terms of basic mineralogy.

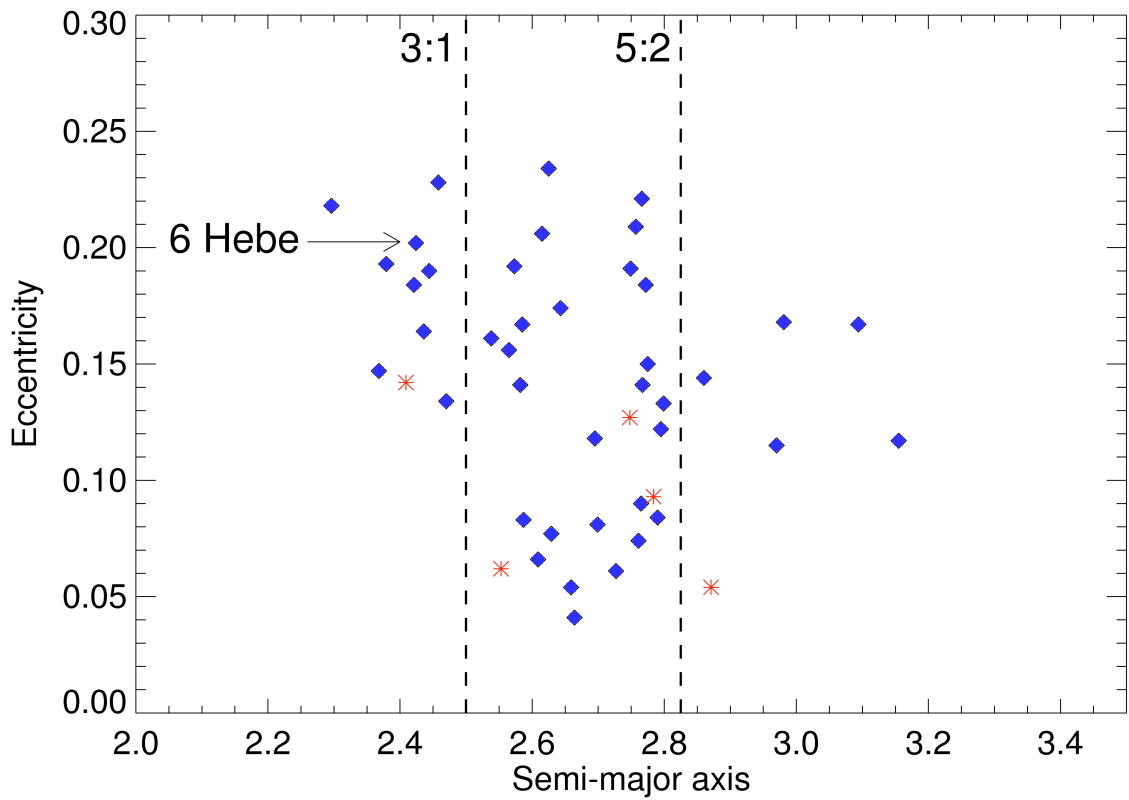

**Figure 5: The location of H-like (Hebe-like) bodies in the asteroid belt.** Families are shown in red [Agnia (2.78, 0.09); Koronis (2.87, 0.05); Maria (2.55, 0.06); Massalia (2.41, 0.14); Merxia (2.75, 0.13)]. We also indicate the location of asteroid (6) Hebe. One can see that asteroid Hebe is not the most favorable source region of H chondrites. Many asteroids lie closer to the 3:1 and 5:2 resonances than Hebe.

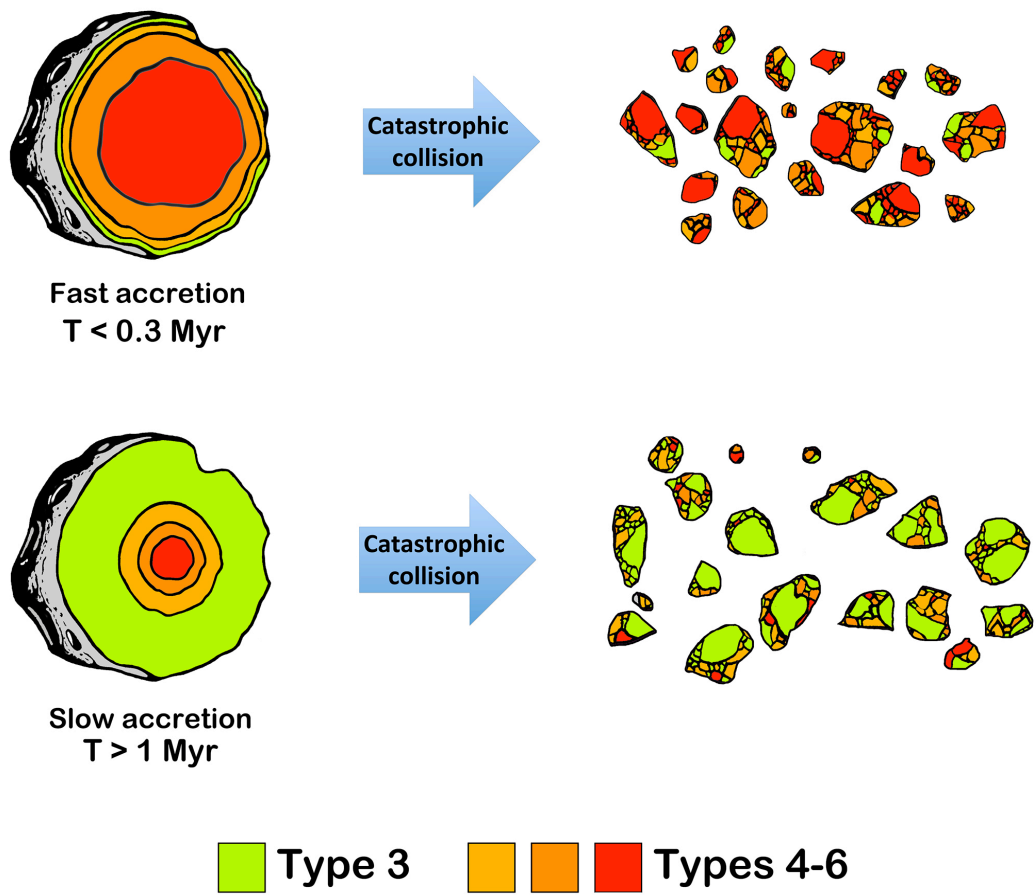

**Figure 6. Asteroid families support fast accretion.** Primordial heating and metamorphism of asteroid interiors, as deduced from meteorite petrology studies implicates the development of an onion skin structure at an early formation stage (left). The duration of accretion will dictate the compositional structure of a primordial OC parent body: 'instantaneous' accretion (top; T<0.3 Myr) will lead to a thin crust of type 3 material while slow accretion will lead to a thick (>10km) external layer of type 3 material (left). Note that instantaneous accretion at a later time would be equivalent in terms of the internal structure of the parent-body. Members of S-type asteroid families (right) have surface compositions compatible with those of type 4-6 OCs. This implies that most of the volume in the primordial parent body has been metamorphosed to high temperature, which is consistent with a short duration of accretion.

The graphics for the onion shell model and the rubble pile were reproduced from original work (McSween & Patchen 1989).

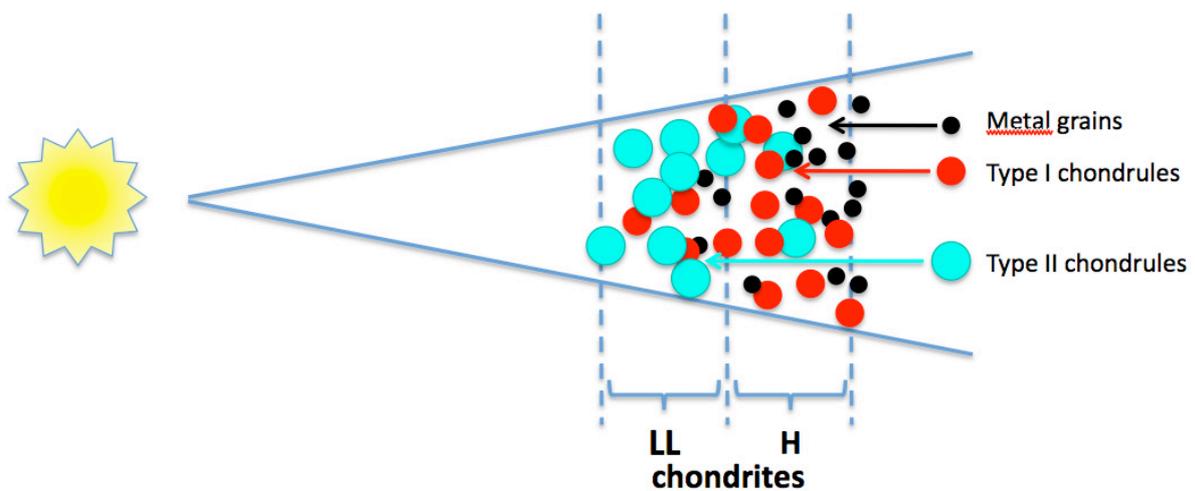

**Figure 7. Formation locations of H and LL chondrites as a consequence of size sorting?** Here, we illustrate the different formation location for both H and LL chondrites, namely LL chondrites formed closer to the Sun than H chondrites. We also add information concerning the building blocks (chondrules, metal grains) of both H and LL chondrites. Chondrules, the main building blocks of OCs, are size-sorted, with the larger (0.6 mm on average), more oxidized type II chondrules more abundant in LL chondrites, while the smaller (0.3 mm on average), more reduced type I chondrules dominate in H chondrites (Zanda et al. 2006).

**Appendix A**

**A.1) Observations and data reduction in the near-infrared**

New data presented here are near-infrared spectral measurements from 0.7 to 2.5 μm obtained using SpeX, the low- to medium-resolution near-IR spectrograph and imager (Rayner et al., 2003), on the 3-meter NASA IRTF located on Mauna Kea, Hawaii. Observing runs were conducted remotely primarily from the Observatory of Paris-Meudon, France. The spectrograph SpeX, combined with a 0.8 x 15 arcsec slit was used in the low-resolution prism mode for acquisition of the spectra in the 0.7-2.5 μm wavelength range. In order to monitor the high luminosity and variability of the sky in the near-IR, the telescope was moved along the slit during the acquisition of the data so as to obtain a sequence of spectra located at two different positions (A and B) on the array. These paired observations provided near-simultaneous sky and detector bias measurements. Objects and standard stars were observed near the meridian to minimize their differences in airmass and match their parallactic angle to the fixed N/S alignment of the slit. Our primary solar analog standard stars were 16 Cyg B and Hyades 64. Additional solar analog stars with comparable spectral characteristics were utilized around the sky. Two to three sets of eight spectra per set were taken for each object, with each with exposures typically being 120 s.

Finally, reduction was performed using a combination of routines within the Image Reduction and Analysis Facility (IRAF) and Interactive Data Language (IDL). See DeMeo et al. (2009) for a full description of the procedure.

**A.2) Robustness of our radiative transfer model**

We applied a radiative transfer model to analyse and compare the mineralogies for both asteroid and meteorite samples. We first applied this model to the ordinary chondrite meteorite spectra to (1) constrain the composition of ordinary chondrites relative to petrographic measurements; and (2) to perform a direct comparison between these meteorites and their possible asteroid parent bodies.

We already applied this model to EOCs in the past (Vernazza et al. 2008). At that time, we showed the strength of our method by demonstrating that our model-fit average ratios are within 3% of their corresponding value derived from direct laboratory analysis (Hutchison 2004).

Recently, the robustness of the method – namely its ability to link a specific meteorite to a given asteroid - has been validated by the samples brought back from the asteroid Itokawa by the Hayabusa mission (Nakamura et al. 2011). In 2008, our technique had successfully predicted that the asteroid Itokawa must have a composition similar to LL chondrites (Vernazza et al. 2008).

Last, it is important to note that our model is applied in a totally consistent way to both meteorites and asteroids, which allows us to be extremely confident in any measured difference.

**A.3) Composition of UOCs based on their spectral properties**

We applied the radiative transfer model to the newly acquired spectra of UOCs in order to constrain the relative abundance of the two main minerals present in these meteorites (olivine, orthopyroxene, see Appendix Table 2). We found that for the lowest petrologic types (<3.5), H, L and LL meteorites have a similar ol/(ol + low-Ca px) ratio (>0.65). For petrologic types greater than 3.6, the compositional trend evolves gradually towards the more familiar compositional spread seen/known for EOCs (Dunn et al. 2010a, Dunn et al. 2010b, Jarosewich 1990): on average 58.8 ± 4.5% for Hs; 64.2 ± 6.8% for Ls; and 75.1 ± 4.5% for LLs (Vernazza et al., 2008).

It is interesting to note that our results are supported by previous independent laboratory measurements that utilized Mossbauer spectroscopy and X-ray diffraction to quantify the modal mineralogy of unequilibrated ordinary chondrites (UOCs). As in our case, these authors (Menzies et al. 2005) noticed a broad decrease in the olivine/pyroxene ratio in the early stages of equilibration, suggesting reduction in the least equilibrated ordinary chondrites.

These variations can possibly be understood using the bulk chemical data of UOCs (Jarosewich 1990) and were already noted by McSween and Labotka (1993) based on the same set of data. We believe these data to indicate (Appendix Figure 4) that C-induced reduction in the early stages of metamorphism provokes oxygen loss and metal formation, leading to the formation of pyroxene at the expense of olivine by the following reaction: $FeMgSiO_4 + C \rightarrow CO + Fe_{met} + MgSiO_3$. This early reduction effect is more pronounced amongst Hs, which contained less oxygen and more C in their starting material, thus offering a possible explanation for the strong compositional divergence with increasing metamorphism with respect to the L and LL groups.

**A.4) Bimodal compositional distribution of main-belt S-type asteroids.**

The dip test (mentioned in the manuscript) is actually not a direct proof of a bimodal distribution. The dip test is a proof against unimodality. A direct test of bimodality (Schilling et al. 2002) says that a mixture of two normal distributions with similar variability cannot be bimodal unless their means differ by more than the sum of their standard deviations. In our case, this criterion is largely fulfilled (see Appendix Figure 5).

**A.5) H chondrite parent bodies: correlation between size and composition**

We observe a correlation between the composition and the asteroid size within the H peak (see Appendix Figure 8). As detailed in section 4.2, for all asteroids in the left peak (H-like bodies) we are seeing their exposed interiors. As such, their surfaces give us a first order representation (sort of insight into the composition of the external layers down to a depth of 30km in a D=200km body; see Ciesla et al. 2013) of the composition of their interior. The observed correlation may thus indicate a variation of the internal composition of H-like bodies as a function of their initial sizes. In particular, the correlation may indicate that smaller H-like bodies formed with a more pyroxene-rich interior than larger H-like bodies (the data for families are in agreement with this hypothesis with Koronis members being on average more olivine-rich than Agnia or Merxia ones). In the context of the H4-H6 sequence (H4<H5<H6 in terms of ol/(ol + low-Ca px) ration; see Dunn et al. 2010a,b) this may imply that smaller bodies were formed with a higher fraction of H4 material while larger bodies formed with more H5 and/or H6 material. Thermal models (e.g. Henke et al. 2012a,b; Henke et al. 2013)) could directly test this hypothesis.

**A.6) H parent bodies smaller than LL ones**

We computed the average size of asteroids in the two peaks in order to trace any initial difference in terms of mean size between the H and LL parent bodies. We selected only asteroids with D>100km as smaller ones are likely the result of the collisional evolution over the course of the history of the Solar System. Note that it is not implied that all D<100km asteroids are the results of collisional disruptions; many D=[70-100]km asteroids may indeed be primordial. In any case, the current size distribution for D>100km gives us a good indication of primordial differences in mean sizes between H and LL bodies.

We found that the average size within the H (left) peak is D = 132±6 km while the average size in the LL (right) peak is D = 160±10 km. Our results imply that LL bodies were born, on average, bigger than H ones. Note that if we measure the average size over a wider range (for all objects with D>60km), this result remains unchanged (D=94km for the H peak and D=119km for the LL peak).

**Appendix Figures**

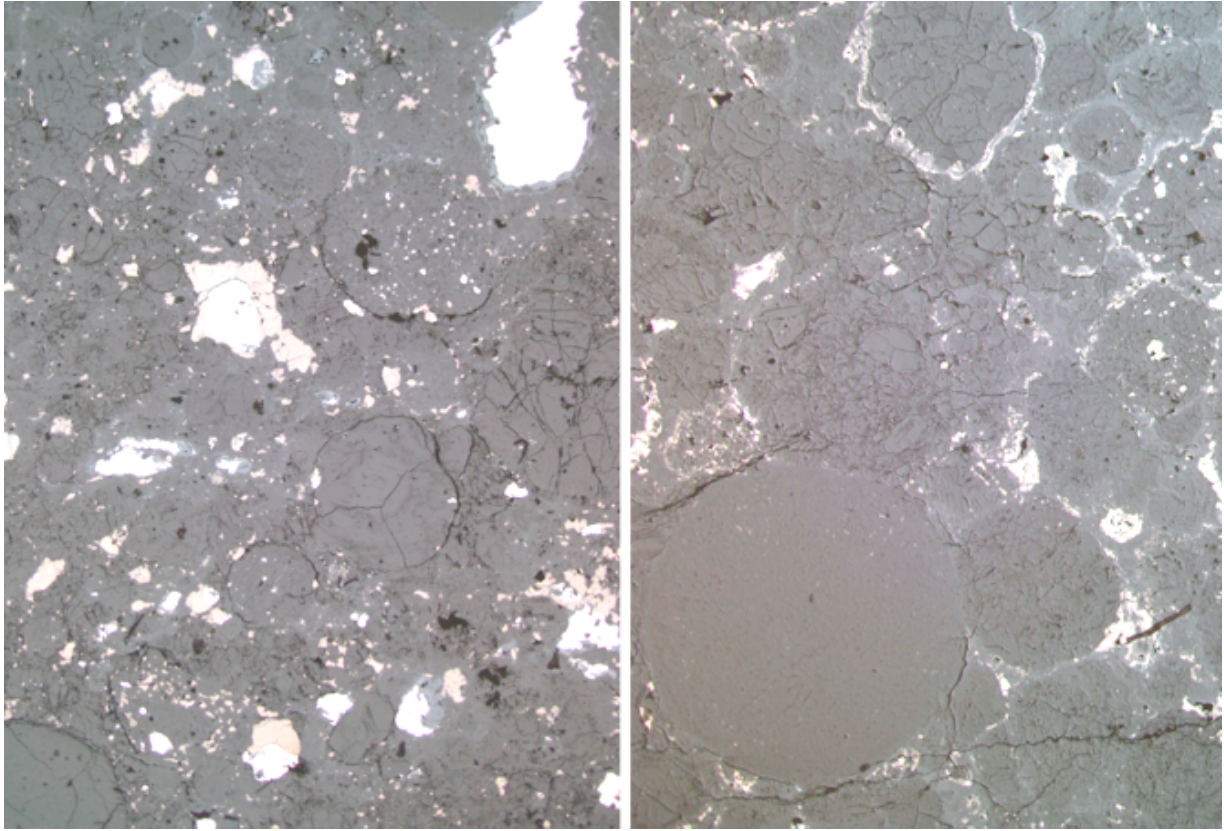

**Figure A.1**: Reflected light images of the Grady 1937 H3 chondrite and of the Krymka LL3 chondrite. Field of view: 4x5.5 mm. Opaque phases: metal (white) and iron sulfide (beige) are mostly located between (or around) the chondrules. The H chondrite comprises much more metal and its chondrules are smaller (mean diameter 300µm – (Scott et al. 1996)) than those in the LL chondrite (mean diameter 900µm – (Scott et al. 1996)).

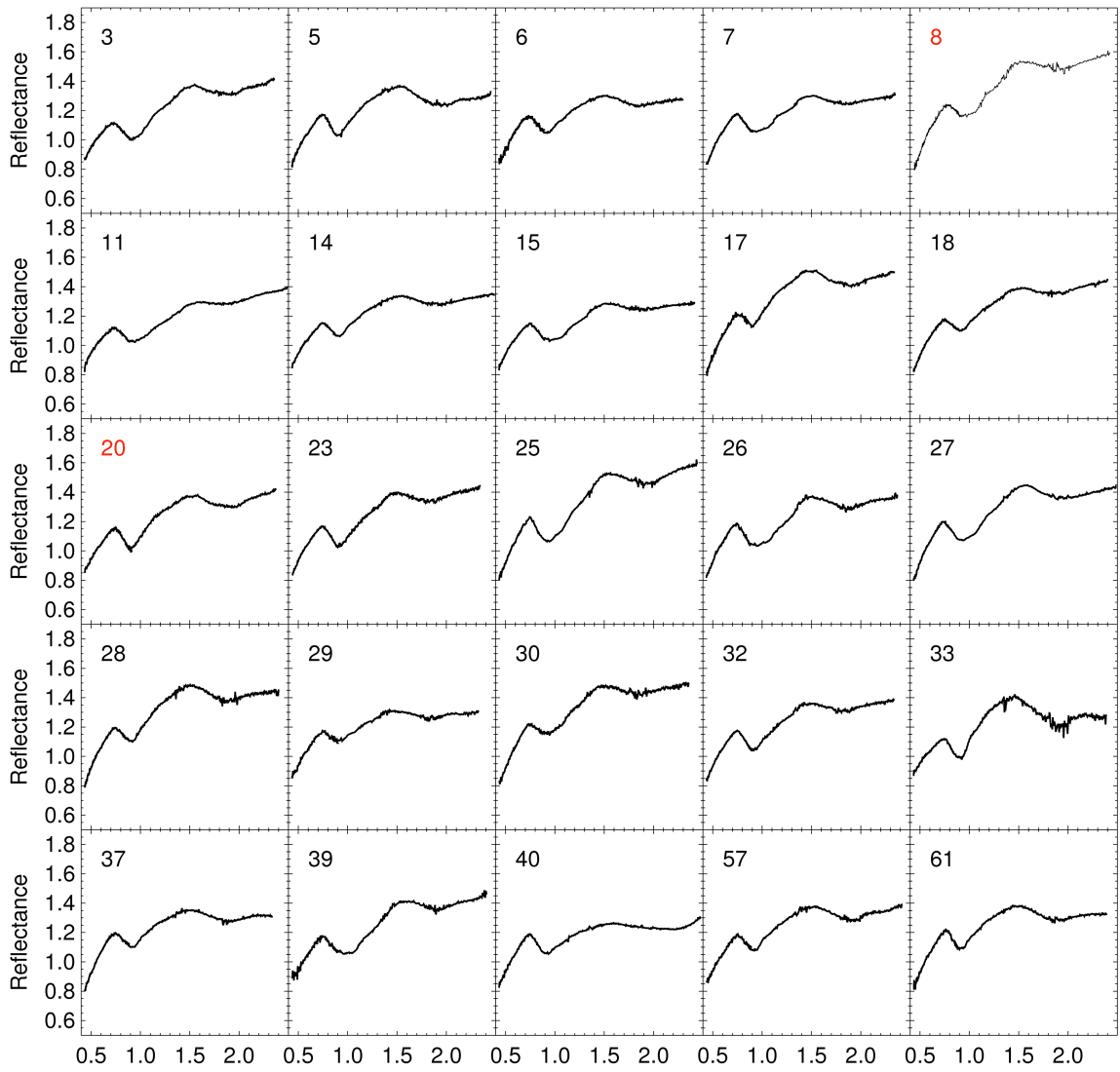

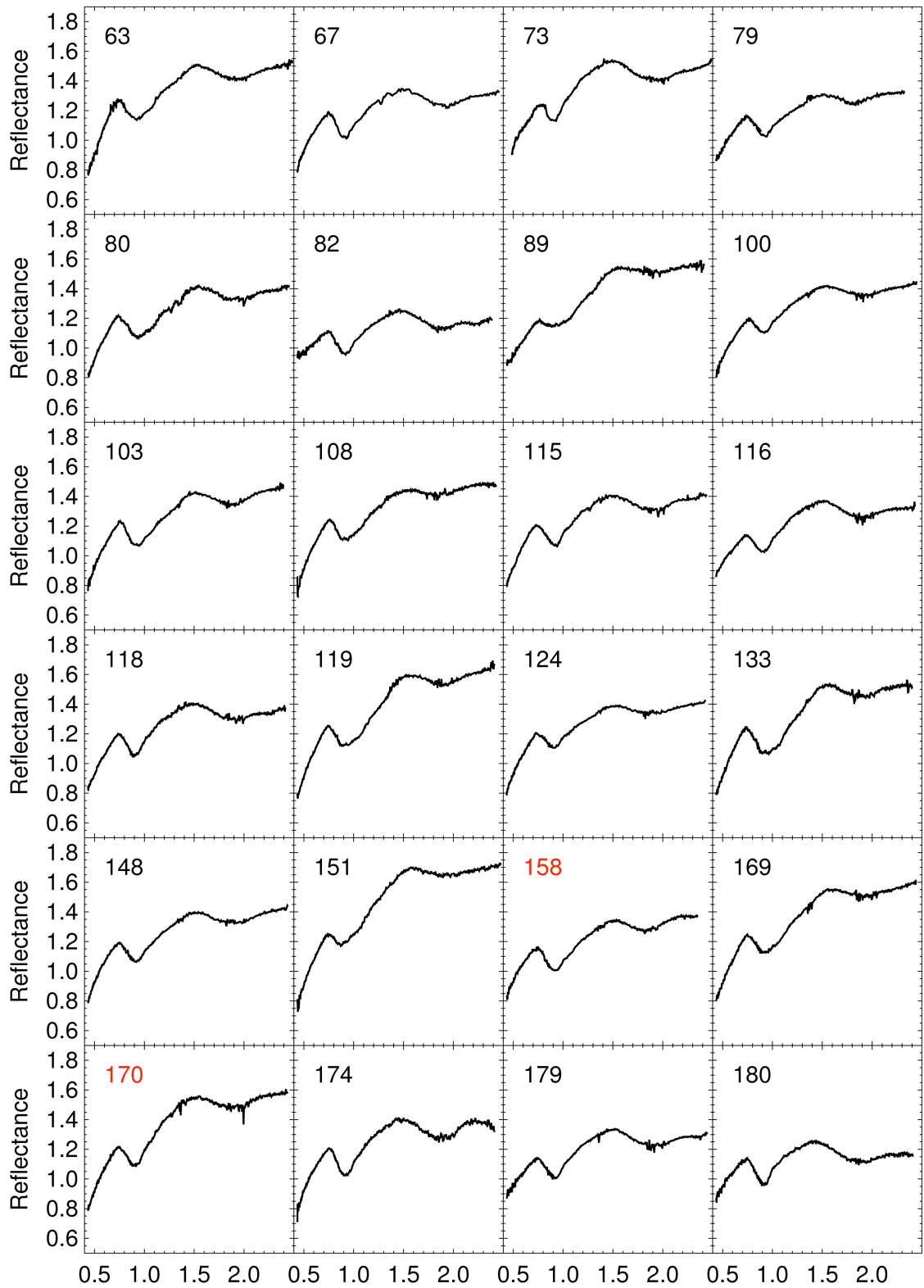

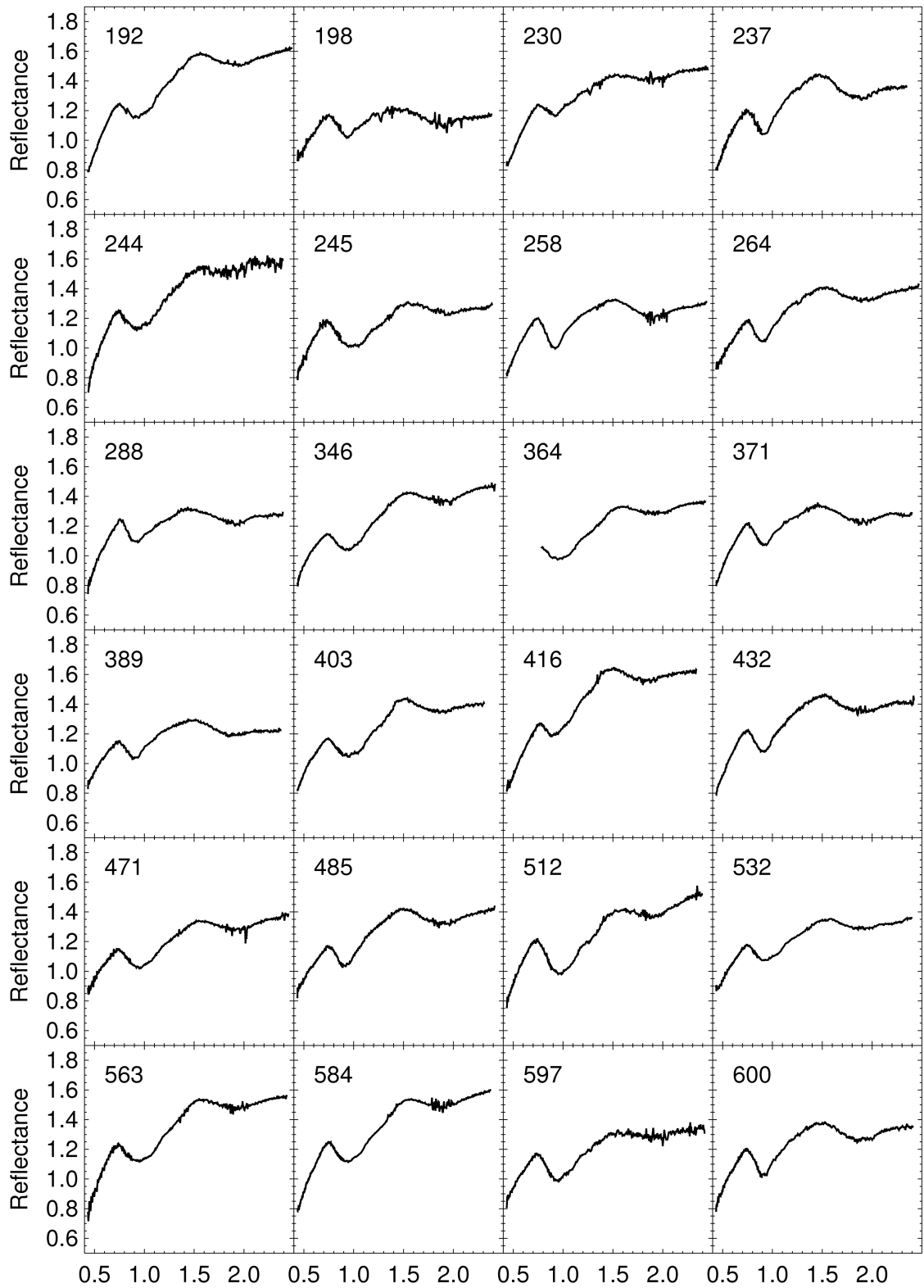

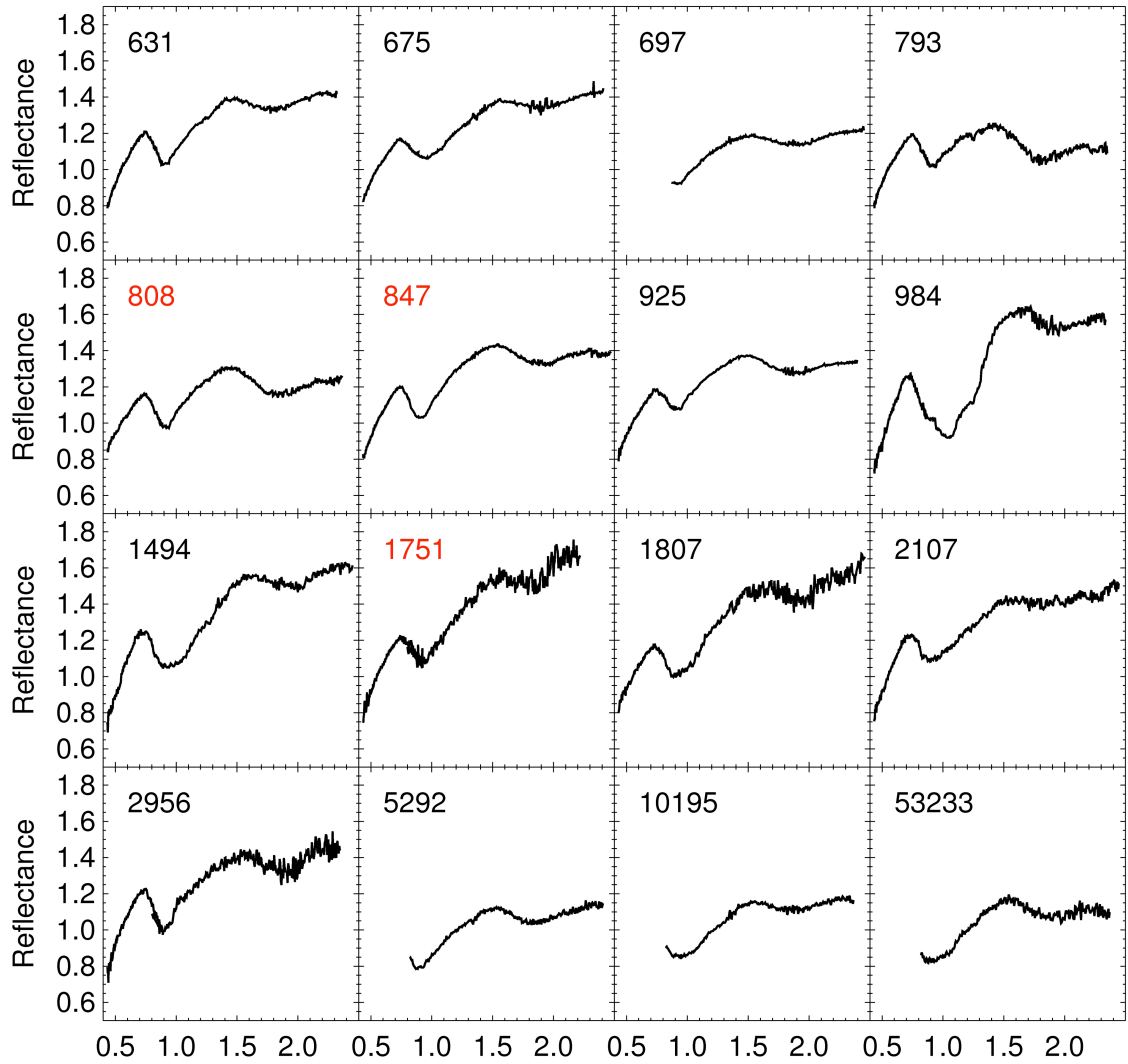

**Figure A.2**: Plots of the final reduced NIR spectra of S-type asteroids presented in this paper combined with available visible wavelength spectra (Bus 1999). In red, we highlight one typical spectrum for each family (8 for Flora, 15 for Eunomia, 20 for Massalia, 158 for Koronis, 170 for Maria, 808 for Merxia, 847 for Agnia, 1751 for Gefion).

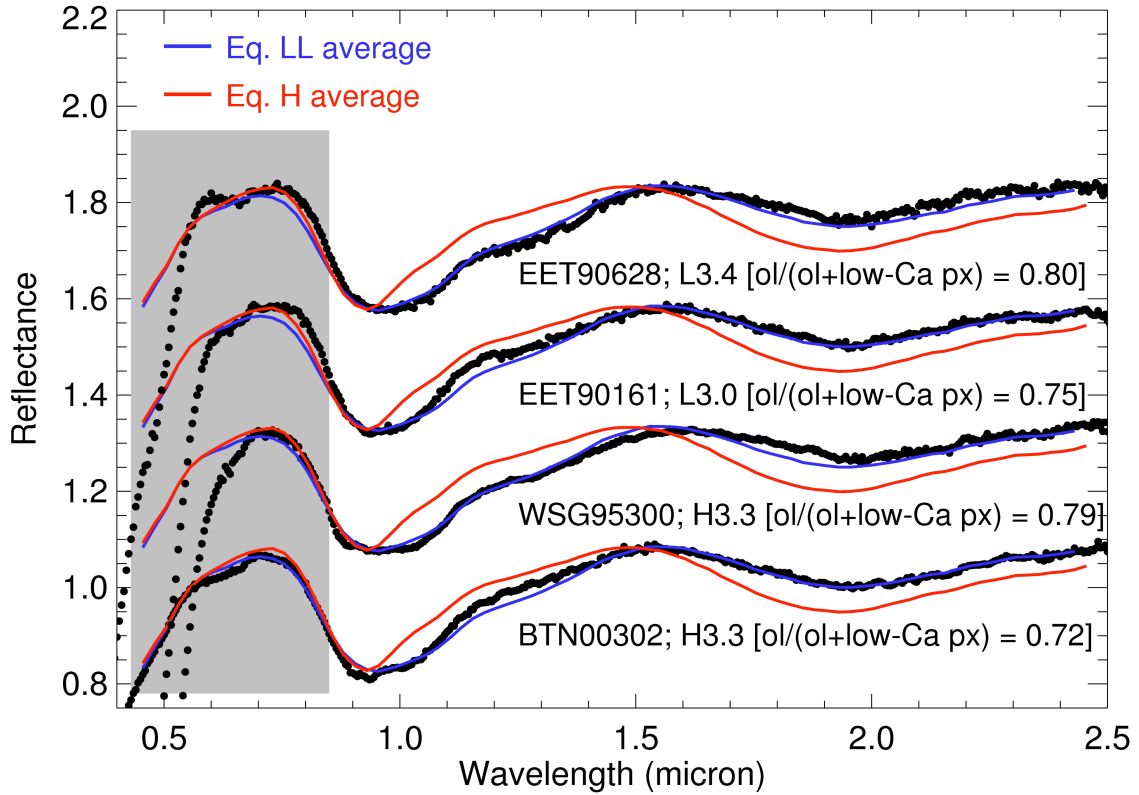

**Figure A.3**: Comparison of the spectral properties of type 3.0-3.4 UOCs with those of EOCs. It appears that type 3.0-3.4 UOCs have, surprisingly, the same spectral properties as equilibrated LL chondrites. This figure also highlights that UOCs are highly affected by terrestrial weathering in the visible domain (grey zone). The spectral effect of this weathering is well known (Gooding, 1982): it causes a drop of the reflectance in the visible and in the UV. As a consequence, we modeled the composition of the samples over the near-IR range only.

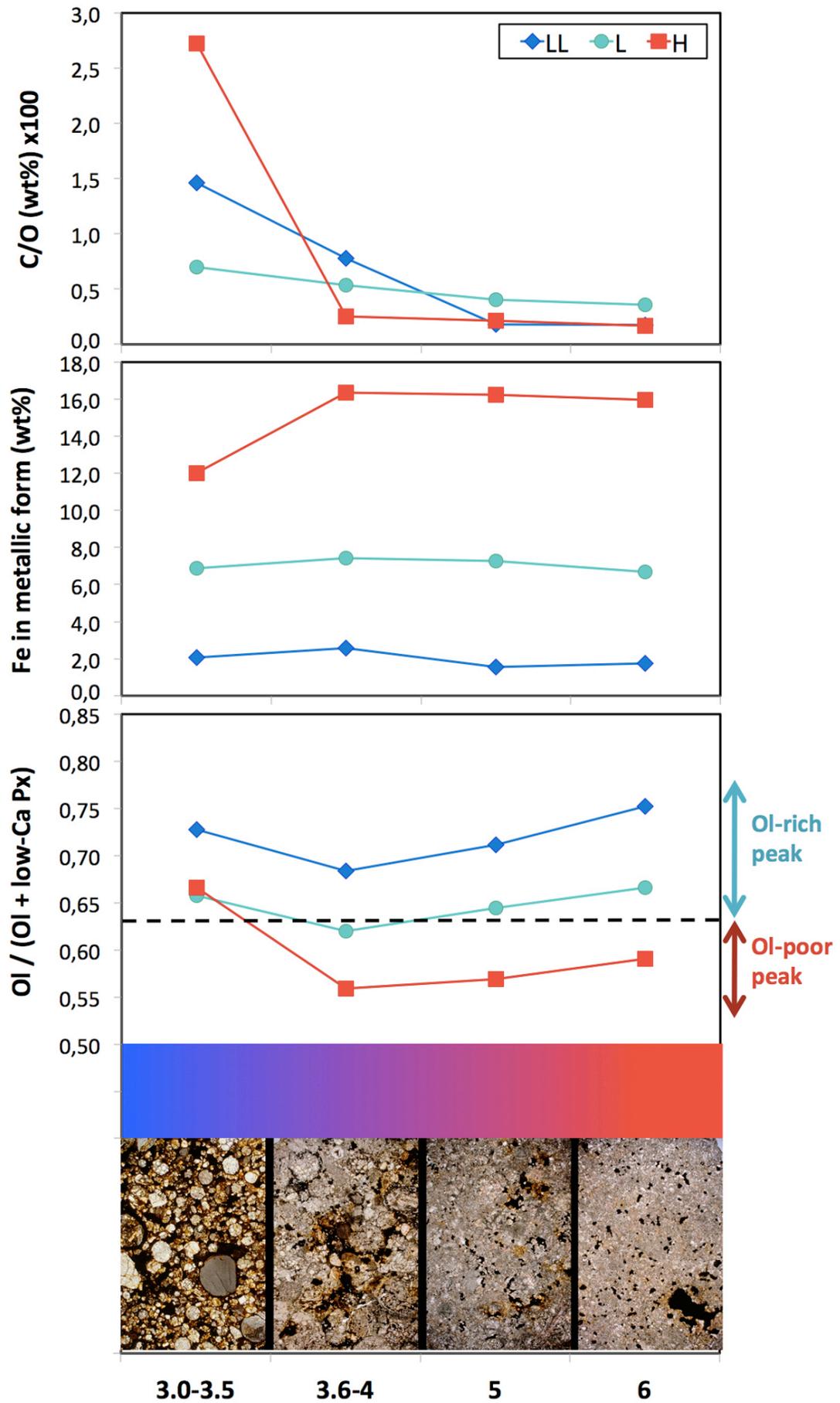

**Figure A.4**: Compositional and textural evolution as a function of petrologic type. (bottom) All images are 4.7 x 7 mm. Chondrites are characterized by the presence of chondrules, sub-millimeter spherules melted in the early solar system by high temperature events of a still unknown nature (Zanda 2004). Chondrules comprise over 80% of the rock in the least metamorphosed ordinary chondrites (lower left picture) and they are separated by matrix, a fine-grained low-temperature material. Metamorphic heating in parent asteroid results in chemical and textural changes, the most spectacular of which is the progressive disappearance of chondrules and matrix, shown here for L chondrites ranging from petrologic type 3 (least heated) to 6 (most heated). Chondrites with petrologic types ≥4 are often referred to as "equilibrated" because their minerals have achieved chemical equilibrium though heat-induced elemental diffusion.

(top) The 3 diagrams are based on a consistent data set of bulk chemical analyses of chondrites (Jarosewich 1990). The top panel shows the evolution of the C/O ratio with petrologic type, and the middle one the abundance of Fe in metallic form. The bottom panel shows the evolution of the ol/(ol + low-Ca px) ratio with petrologic type. We calculated normative abundances of olivine and orthopyroxene from bulk chemical analyses (Jarosewich 1990). We also calculated the ferromagnesian silicate part of the bulk composition for the least equilibrated chondrites and used their C contents to reduce this silicate to verify the trends in ol/(ol + low-Ca px) through the more equilibrated chondrites.

This early reduction effect, already noted by Menzies et al. (2005), is more pronounced amongst Hs, which contained less oxygen and more C in their starting material. It results in a rapid increase in metal abundance from type 3s to type 4s, which is followed by a slight decrease from type 4 to type 6, due to subsequent oxidation (McSween and Labotka, 1993; Gastineau-Lyons et al., 2002; Dunn et al., 2010; this work). Reduction of Fe from olivine in the least equilibrated chondrites makes Si available to create more pyroxene and thus decreases the ol/(ol + low-Ca px) ratio as shown in the lower panel, a trend that is reversed in higher petrologic types.

This diagram shows that, whereas all type 3s have fairly similar ol/(ol + low-Ca px) ratios in all the groups, H4, H5 and H6 diverge by being significantly more depleted in olivine. The position of the dip/gap observed in our asteroidal composition distribution (histogram) is also shown here: metamorphosed H chondrites fall below it, while metamorphosed LL chondrites do not, allowing us to unambiguously equate the olivine-poor peak to metamorphosed H chondrites. Note that metamorphosed L chondrites straddle the gap between the two peaks. The very presence of this gap (instead of a continuous distribution) indicates that asteroids corresponding to metamorphosed L chondrites must now be very rare in the main belt. Note that for LL chondrites, the chondrites chemical analyses show that the bulk of the reduction takes place at lower temperatures as for Ls and Hs, specifically between Semarkona (LL3.00) and Krymka (LL3.2), so we chose to artificially aggregate LL3.2-LL3.3.5 with petrologic types 3.6-4 on this diagram, in order to reveal this reduction. If these objects had been kept within the 3.0-3.5 bin, then the ol/(ol+px) ratio in that bin would be lower and the ratio in the 3.6-4 bin would be higher, resulting in a continuously increasing trend.

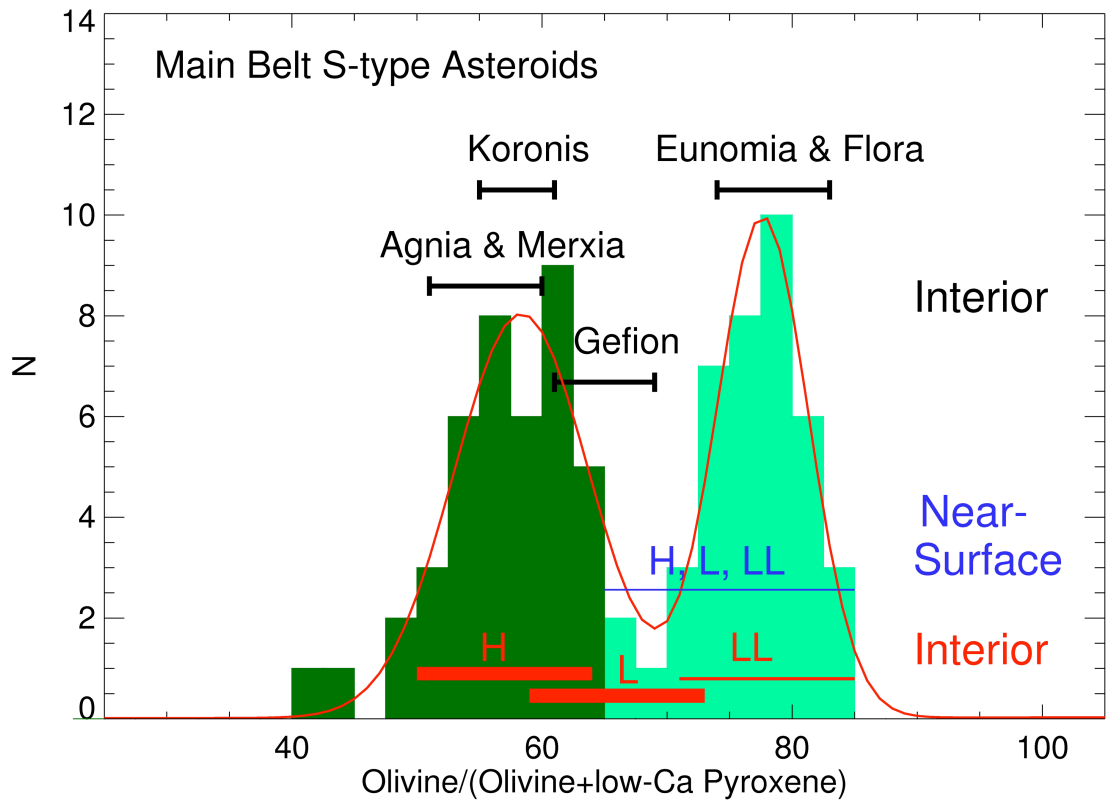

**Figure A.5**: Bimodal compositional distribution of main-belt S-type asteroids. (Objects belonging to collisional families are not included in the histogram counts.) Here we show our data best fitted by two gaussians. The parameters (mean and standard deviation) of the two gaussians are as following: 58.35±5.41 and 77.64±3.66. Following the definition of a bimodal distribution (Schilling et al. 2002), a separation in means of at least 1.16*(5.41+3.66)= 10.52 is necessary for bimodality. In our case, the separation in means is 19.29, that is more than a factor of two than the sum (9.07) of their standard deviations! This demonstrates that our distribution is bimodal.

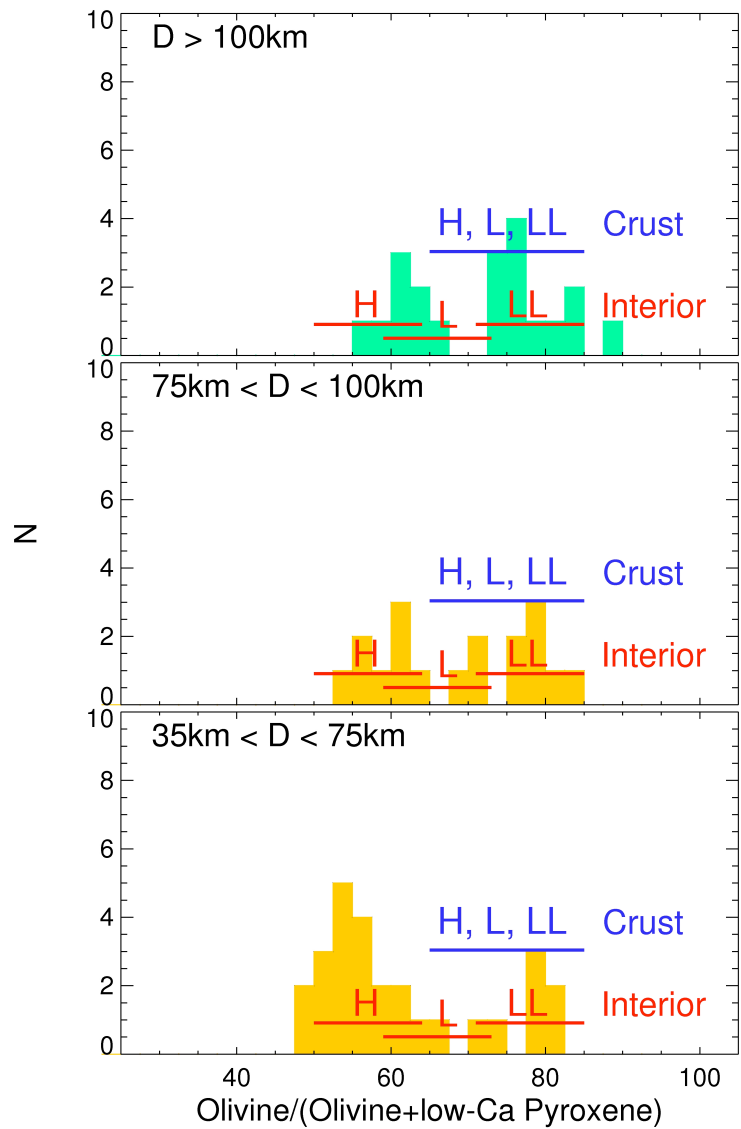

**Figure A.6**: Compositional distribution of main-belt S-type asteroids as a function of their size. The bimodality is seen at large sizes implying that the bimodality is a primordial signature and not the consequence of the collisional evolution. In the LL (right peak), the excess of large bodies (D>75km) with respect to smaller ones (35km<D<75km) suggests the idea that D>~100km bodies are primordial. In the H (left) peak, one can see the correlation between the size and the composition (see Appendix Figure 8).

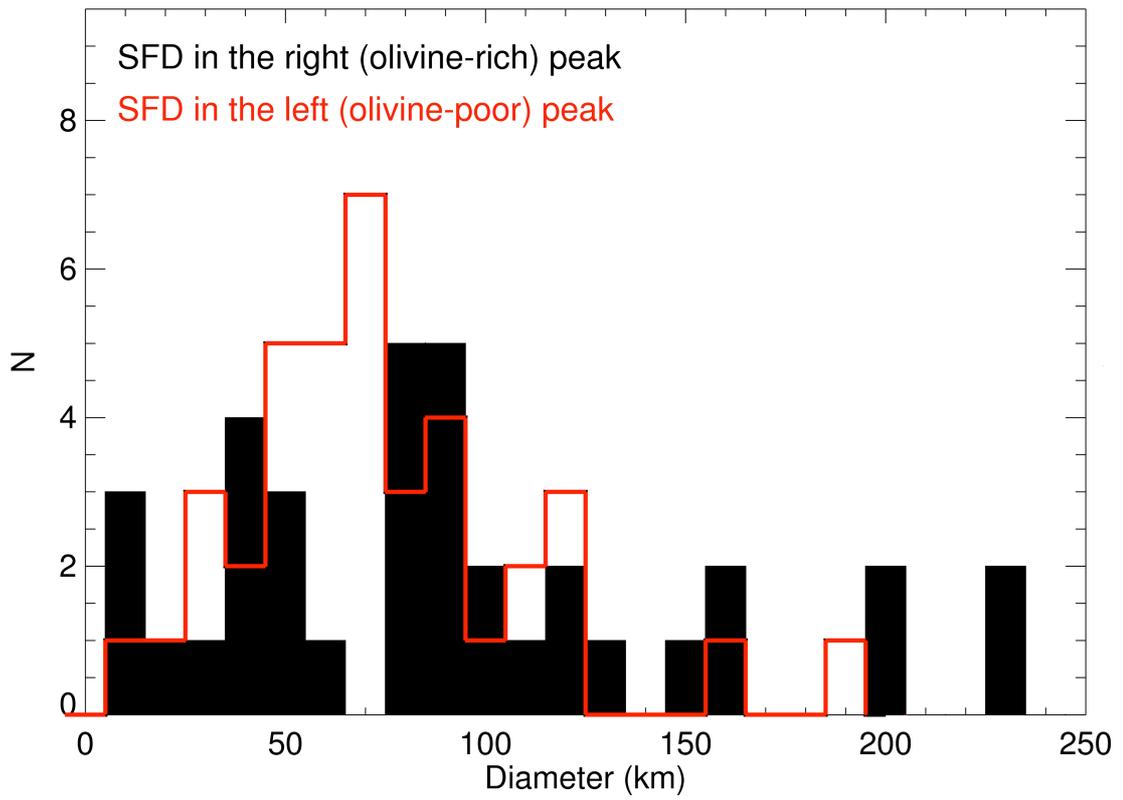

**Figure A.7**: Size frequency distribution of the objects in the left (red) and right (black, filled) peaks. The diameters were taken from Masiero et al. (2011). Objects in the olivine-poor peak appear smaller than those in the olivine-rich peak.

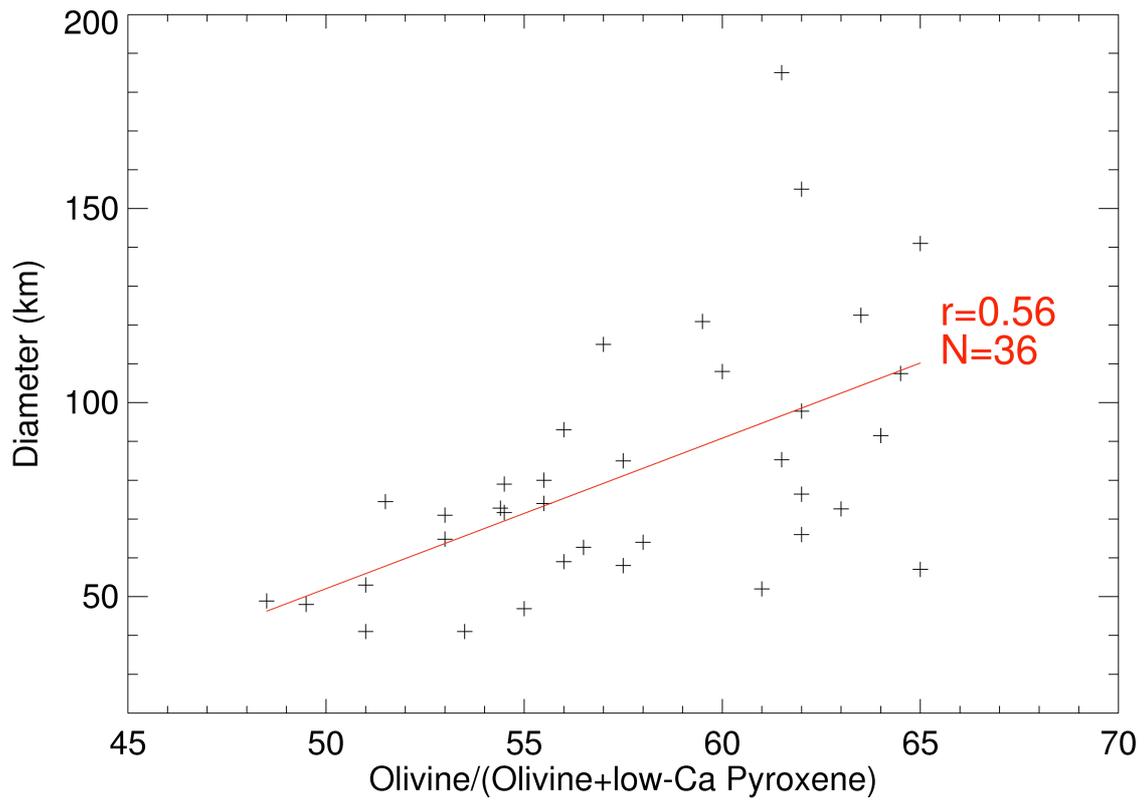

**Figure A.8:** The relationship between the size of olivine-poor (that is, ol/(ol + low-Ca px)=45–65%) S-type asteroids and their composition. We observe a linear relation between the size and composition with a correlation coefficient of **r**=0.56 (**N**=36), a >99.9% confidence level that this correlation is not random.

**Table A.1: Asteroid Composition**

| Asteroid | Data Reference[a] (NIR) | [ol/(ol + low-Ca px)] % | Semi-major axis (AU) | D (km)[b] | Albedo[b] |
|---|---|---|---|---|---|
| 3 | 6 Jan 2011 | 72.5 | 2.671 | 234 | 0.24 |
| 5 | 20 Feb 2004 | 57 | 2.573 | 115 | 0.25 |
| 6 | 15 Jan 2008 | 61.5 | 2.424 | 185 | 0.27 |
| 7 | 20 Feb 2004 | 81 | 2.385 | 200 | 0.28 |
| 11 | 13 Nov 2005 | 79 | 2.452 | 159 | 0.16 |
| 14 | 1 Oct 2003 | 62 | 2.585 | 155 | 0.22 |
| 17 | 6 Jan 2011 | 56 | 2.47 | 93 | 0.16 |
| 18 | 10 Jan 2011 | 65 | 2.296 | 141 | 0.22 |
| 23 | 10 Jan 2011, 6 Jan 2011 | 64.5 | 2.625 | 107.5 | 0.25 |
| 25 | 04 Sep 2010, 13 Oct 2010 | 74.5 | 2.399 | 75 | 0.23 |
| 26 | 11 Jul 2011 | 80.5 | 2.658 | 87 | 0.23 |
| 27 | 22 Sep 2004 | 77 | 2.346 | 118 | 0.20 |
| 28 | 04 Sep 2010 | 59.5 | 2.775 | 120.9 | 0.18 |
| 29 | 12 Jul 2010 | 73 | 2.554 | 227 | 0.16 |
| 30 | 12 Jul 2010 | 78 | 2.365 | 98.4 | 0.17 |
| 32 | 3 Dec 2008 | 61.5 | 2.587 | 85.3 | 0.23 |
| 33 | 04 Sep 2010 | 43.5 | 2.865 |  | 0.24 |
| 37 | 04 Sep 2010, 13 Oct 2010 | 60 | 2.643 | 108 | 0.18 |
| 39 | 11 Jul 2010 | 87.5 | 2.769 | 163 | 0.24 |
| 40 | 16 Oct 2004 | 72.5 | 2.267 | 120 | 0.20 |
| 57 | 11 Jul 2010 | 63.5 | 3.155 | 122.5 | 0.18 |
| 61 | 04 Sep 2010, 14 Oct 2010 | 57.5 | 2.981 | 85 | 0.21 |
| 63 | 30 Sep 2003 | 75 | 2.395 | 103 | 0.16 |
| 67 | 16 Jun 2004 | 56.5 | 2.421 | 62.7 | 0.22 |
| 73 | 16 Oct 2003 | 48.5 | 2.664 | 48.9 | 0.19 |
| 79 | 30 Oct 2008 | 63 | 2.444 | 72.6 | 0.22 |
| 80 | 03 Sep 2010 | 78.5 | 2.295 | 79 | 0.18 |
| 82 | 11 Jul 2010 | 55.5 | 2.766 | 74 | 0.14 |
| 89 | 10 Jan 2011 | 82.5 | 2.550 | 148 | 0.18 |
| 100 | 04 Sep 2010 | 64 | 3.094 | 91.5 | 0.18 |
| 101 | 22 Dec 2006 | 62 | 2.582 | 66 | 0.19 |
| 103 | 11 Jul 2010 | 72 | 2.703 | 85 | 0.21 |
| 108 | 12 Jul 2010 | 68.5 | 3.249 | 75.5 | 0.15 |
| 115 | 04 Sep 2010 | 55.5 | 2.379 | 80 | 0.25 |
| 116 | 04 Sep 2010 | 54.5 | 2.767 | 71.7 | 0.26 |
| 118 | 03 Sep 2010, 25 Sep 2011 | 51 | 2.436 | 53 | 0.14 |
| 119 | 11 Jul 2010 | 78 | 2.582 | 61 | 0.20 |
| 123 | 04 Sep 2010 | 49.5 | 2.695 | 48 | 0.21 |
| 124 | 10 Jan 2011 | 62 | 2.629 | 76.4 | 0.17 |
| 133 | 10 Jan 2011 | 77.5 | 3.062 | 80.5 | 0.18 |
| 148 | 10 Jan 2011 | 62 | 2.772 | 97.8 | 0.16 |
| 151 | 13 Nov 2005 | 70.5 | 2.593 | 41 | 0.21 |
| 169 | 04 Sep 2010 | 77.5 | 2.358 | 38.5 | 0.18 |
| 174 | 10 Jan 2011 | 51.5 | 2.86 | 74.5 | 0.13 |
| 179 | 04 Sep 2010 | 54.4 | 2.970 | 72.8 | 0.18 |
| 180 | 12 Jul 2010 | 41.5 | 2.722 | 24.35 | 0.23 |
| 192 | 30 Apr 2006 | 77 | 2.404 | 93 | 0.29 |
| 198 | 05 Sep 2010 | 65 | 2.458 | 57 | 0.26 |
| 230 | 03 Sep 2010 | 82.5 | 2.382 | 109 | 0.17 |
| 237 | 10 Sep 2010, 27 Oct 2010 | 51 | 2.761 | 41 | 0.21 |
| 244 | 13 Apr 2008 | 77.5 | 2.174 | 11.4 | 0.18 |
| 245 | 31 Oct 2008 | 84 | 3.101 | 78 | 0.22 |
| 258 | 04 Sep 2010 | 53 | 2.615 | 64.8 | 0.17 |
| 264 | 19 May 2005 | 53 | 2.799 | 71 | 0.15 |
| 288 | 12 Jul 2010 | 63 | 2.757 | 32 | 0.17 |
| 346 | 10 Jan 2011 | 75 | 2.797 | 92 | 0.29 |
| 364 | 20 July 2007, 2 Oct 2007 | 82 | 2.22 | 28 | 0.26 |
| 371 | 12 Jul 2010 | 56 | 2.727 | 59 | 0.16 |
| 389 | 04 Sep 2010, 14 Oct 2010 | 54.5 | 2.609 | 79 | 0.20 |
| 403 | 6 Sep 2010 | 80.5 | 2.81 | 49.5 | 0.17 |
| 416 | 7 Sep 2010 | 71 | 2.789 | 85.5 | 0.17 |
| 432 | 10 Jan 2011 | 55 | 2.368 | 46.9 | 0.23 |
| 471 | 04 Sep 2010 | 77 | 2.888 | 134 | 0.20 |
| 485 | 10 Jan 2011 | 58 | 2.749 | 64 | 0.21 |
| 512 | 10 Sep 2010 | 80.5 | 2.189 | 23 | 0.18 |
| 532 | 3 Dec 2008 | 75 | 2.770 | 203 | 0.20 |
| 563 | 04 Sep 2010 | 82 | 2.713 | 53.5 | 0.25 |
| 584 | 11 Nov 2007 | 79.5 | 2.373 | 49 | 0.24 |
| 597 | 10 Sep 2010 | 73 | 2.673 | 36 | 0.24 |
| 600 | 10 Sep 2010, 22 Aug 2011 | 56 | 2.659 | 28.3 | 0.18 |
| 631 | 10 Sep 2010, 22 Aug 2011 | 61 | 2.790 | 52 | 0.21 |
| 675 | 10 Sep 2010 | 79 | 2.769 | | |
| 695 | 01 Nov 2010 | 53.5 | 2.538 | 41 | 0.25 |
| 793 | 12 Jul 2010 | 58 | 2.795 | 30 | 0.15 |
| 925 | 05 Sep 2010, 25 Sep 2011 | 57.5 | 2.699 | 58 | 0.25 |
| 984 | 12 Jul 2010 | 93.5 | 2.803 | 35.5 | 0.38 |
| 1494 | 25 Oct 2006 | 79 | 2.190 | | |
| 1807 | 15 Sep 2004 | 73 | 2.226 | 9.5 | 0.29 |
| 2107 | 29 Jan 2006 | 73 | 2.626 | | |
| 2956 | 11 Nov 2007 | 58.5 | 2.765 | 9.5 | 0.29 |
| 5292 | 24 Oct 2006 | 61.5 | 2.565 | | |
| 10195 | 20 Nov 2006 | 75 | 2.885 | 11 | 0.19 |
| 53233 | 21 Jan 2007 | 75 | 2.380 | | |

[a]For observations reported here, we give the observation date (UT). All near-infrared (NIR) data were obtained using the NASA IRTF at Mauna Kea, Hawaii.

Note: The signal to noise ratio for most asteroid spectra is generally above 100, and above 50 for all objects. This, combined with the reproducibility of repeated measurements yields an accuracy of our ol/(ol + low-Ca px) ratio within a few % (<5).

[b]The albedo and diameters were taken from IRAS and/or WISE (Masiero et al. 2011).

Table A.2: Composition of Unequilibrated Ordinary Chondrites derived from their spectral properties

| Meteorite | Class | [ol/(ol+opx] % |
|---|---|---|
| WSG 95300 | H3.3 | 79 |
| BTN 00301 | H3.3 | 73.5 |
| BTN 00302 | H3.3 | 72 |
| LAR 04382 | H3.4 | 72 |
| MET 00506 | H3.4 | 70.5 |
| MET 00607 | H3.4 | 69 |
| EET 83248 | H3.5 | 58 |
| MAC 88174 | H3.5 | 68.5 |
| EET 83267 | H3.6 | 54.5 |
| ALH 85121 | H3.7 | 53 |
| ALHA77299 | H3.7 | 61.5 |
| GRA 95208 | H3.7 | 65 |
| DOM 03219 | H3.8 | 68 |
| GRA 98023 | H3.8 | 49 |
| RKPA80205 | H3.8 | 56.5 |
| MET 01182 | H3.8 | 68 |
| EET 90161 | L3.05 | 75 |
| QUE 97008 | L3.05 | 80 |
| LEW 86018 | L3.1 | 60.5 |
| MET 96503 | L3.1 | 69 |
| GRO 95502 | L3.2 | 65 |
| GRO 95544 | L3.2 | 66 |
| GRO 95536 | L3.3 | 68 |
| LEW 86127 | L3.3 | 66 |
| MAC 88199 | L3.3 | 68.5 |
| EET 90628 | L3.4 | 80.5 |
| LEW 85339 | L3.4 | 73 |
| LEW 86505 | L3.4 | 73.5 |
| PRE 95401 | L3.4 | 69 |
| GRO 95504 | L3.5 | 67 |
| GRO 95542 | L3.5 | 65.5 |
| GRO 95550 | L3.5 | 67.5 |
| LEW 87284 | L3.6 | 70 |
| ALH 85070 | L3.6 | 70 |
| GRO 06054 | L3.6 | 81 |
| MET 00489 | L3.6 | 76.5 |
| ALH 85155 | L3.7 | 72 |
| ALH 84086 | L3.8 | 66 |
| ALH 84120 | L3.8 | 67 |
| ALH 85045 | L3.8 | 72.5 |
| TIL 82408 | LL3.1-3.5 | 75.5 |
| ALHA76004 | LL3.2/3.4 | 68 |
| ALH 83007 | LL3.2/3.5 | 68.5 |
| ALH 83010 | LL3.3 | 72.5 |
| GRO 95658 | LL3.3 | 68 |
| ALH 84126 | LL3.4 | 77 |
| ALHA78119 | LL3.5 | 70 |
| LEW 87254 | LL3.5 | 74 |
| ALHA77278 | LL3.7 | 75.5 |
| EET 83213 | LL3.7 | 74 |
| GRO 95596 | LL3.8 | 72.5 |
| LAR 06469 | LL3.8 | 68.5 |
| LAR 06301 | LL3.8 | 70 |